\documentclass[reprint,prl,superscriptaddress,longbibliography,amsmath,amssymb,aps] {revtex4-1}
  \usepackage{graphicx}% Include figure files
  \usepackage{dcolumn}% Align table columns on decimal point
  \usepackage{bm}% bold math
  \usepackage{afterpage}
  \usepackage{soul}
  \usepackage{float}
  \usepackage{mathrsfs}
  \usepackage[scr=boondoxo,scrscaled=1.05]{mathalfa} %this seems to be a bit more balanced version of mathrsfs, maybe we could use this 
  \usepackage{placeins}
  \usepackage{dsfont}
  \usepackage{xcolor}
  \usepackage{amssymb}
  \usepackage{amsmath}
  \usepackage{amsfonts}
  \usepackage{graphicx}
  \usepackage{dcolumn}
  \usepackage{cleveref}
  
  \crefrangeformat{equation}{eqs. #3(#1)#4--#5(#2)#6}
  
  \newcommand{\avg}[1]{\left< #1 \right>}

  \usepackage{mathtools}
  \DeclarePairedDelimiter\klam{(}{)}
  \newcommand{\dr}[1]{\mathrm{#1}}
  \newcommand{\nexp}[1]{\dr{e}^{#1}}
  \newcommand{\x}{\dr{i}}
  \newcommand{\laplace}{\mathop{{}\bigtriangleup}\nolimits\!}
  \renewcommand{\vec}[1]{\boldsymbol{\mathrm{#1}}}

\begin{document}
  		\title{Dynamics in Systems with Modulated Symmetries}
  		%: \\ A cellular automaton approach}
  %	\title{\emph{Modulated} symmetries and long-time dynamics: \\ A cellular automaton approach}
  	
  	\author{Pablo Sala}
  	\email{These authors contributed equally to this work.}
  	\affiliation{Department of Physics and Institute for Advanced Study, Technical University of Munich, 85748 Garching, Germany}
  	\affiliation{Munich Center for Quantum Science and Technology (MCQST), Schellingstr. 4, D-80799 M{\"u}nchen, Germany}

  	\author{Julius Lehmann}
  	\email{These authors contributed equally to this work.}
  		\affiliation{Department of Physics and Institute for Advanced Study, Technical University of Munich, 85748 Garching, Germany}
  		\affiliation{Munich Center for Quantum Science and Technology (MCQST), Schellingstr. 4, D-80799 M{\"u}nchen, Germany}
  
  		\author{Tibor Rakovszky}
  		\affiliation{Department of Physics, Stanford University, Stanford, CA 94305, USA}
  	
  	\author{Frank Pollmann}
  		\affiliation{Department of Physics and Institute for Advanced Study, Technical University of Munich, 85748 Garching, Germany}
  	\affiliation{Munich Center for Quantum Science and Technology (MCQST), Schellingstr. 4, D-80799 M{\"u}nchen, Germany}

  	\begin{abstract}
     We extend the notions of multipole and subsystem symmetries to more general {\it spatially modulated} symmetries.
     We uncover two instances with exponential and (quasi)-periodic modulations, and provide simple microscopic models in one, two and three dimensions.
     Seeking to understand their effect in the long-time dynamics, we numerically study a stochastic cellular automaton evolution that obeys such symmetries.
     We prove that in one dimension, the periodically modulated symmetries lead to a diffusive scaling of correlations modulated by a finite microscopic momentum.
     %
%     In higher dimension, these symmetries result in exotic forms of sub-diffusive behavior with a rich spatial structure.
    In higher dimensions, these symmetries take the form of lines and surfaces of conserved momenta.
    These give rise to exotic forms of sub-diffusive behavior with a rich spatial structure influenced by lattice-scale features.
     %
     %These features bear similarity to the phenomenon of ``UV/IR-mixing''.
     %
%     Both of these are novel instances \st{featuring} \pab{resembling} the phenomenon of ``UV/IR-mixing''.   
     %
     Exponential modulation, on the other hand, can lead to correlations that are infinitely long-lived at the boundary, while decaying exponentially in the bulk.
     %, resembling the physics of strong zero modes.
    
  	\end{abstract}
  	
  	%\pacs{Valid PACS appear here}
  	\maketitle
  	%\tableofcontents

  	\textit{\textbf{Introduction.}}
  	Unconventional symmetries can give rise to interesting phenomena, such as new equilibrium phases of matter with novel low-energy features~\cite{Chamon_05,Yoshida_2013,Vijay_2015,Vijay_16,Pretko_17,You_2018,you2020fracton,Devakul19} and unusual non-equilibrium properties such as sub-diffusive transport~\cite{Zhang_2020,Gromov_2020, Morningstar_2020,Feldmeier_2020,Iaconis_2019, glorioso2021breakdown} and Hilbert space fragmentation~\cite{Sala_2020,Khemani_2020, Moudgalya_2020}. Two cases that have been investigated extensively in recent years are multipole moment (e.g. dipole) and subsystem symmetries. The latter often exhibit ``UV/IR''-mixing~\cite{Seiberg_2020,gorantla2021lowenergy,you2020fracton,you2021fractonic}: their long-wavelength properties are sensitive to lattice-scale features, leading to distinct field-theoretic and hydrodynamic descriptions. 
  	A key feature of these symmetries is that they fail to commute with spatial translations~\cite{Gromov_2019}; the corresponding conserved quantities are {\it spatially modulated},  $\mathcal{Q}_{\{\alpha_{\vec{r}}\}}=\sum_{\vec{r}} \alpha_{\vec{r}} q_{\vec{r}}$, where $q_{\vec{r}}$ is some local charge at location $\vec{r}$.
  	For multipole conservation %The dipole and its higher-moments 
  	$\alpha_{\vec{r}}$ is a polynomial of the coordinates $r_i=x,y,z$, while for subsystem symmetries $\alpha_{\vec{r}}$ takes non-zero values only on a spatial submanifold.

  	In the present work, we extend this notion to more general cases of $\alpha_{\vec{r}}$. %spatially modulated symmetries, where $\alpha_{\vec{r}}$ become non-polynomial functions of $r_i$.
  	We show that such symmetries appear in some simple, locally interacting systems, and give various examples in both one (1D), two (2D) and three (3D) dimensions.
  	The symmetries we identify come in two flavors. One type is exponentially localized at the boundaries of the system, leading to infinitely long-lived boundary correlations, resembling the physics of {\it strong zero modes}~\cite{Fendley_2012}. 
  	The other type corresponds to the conservation of certain momentum-components of a local observable. 
  	Interestingly, in 2D and 3D we find various models where long-lived modes exist along some closed hypersurfaces in momentum-space; resembling the recently discussed Bose-Luttinger liquids~\cite{you2020fracton,Lake_2021} and the UV/IR-mixing phenomenon~\cite{Seiberg_2020,gorantla2021lowenergy}. 
  	We will discuss the effect of these symmetries on the system's dynamics and show that they lead to unusual features, such as long-lived spatial oscillations in the correlation functions on microscopic scales.

  	\textit{\textbf{Modulated symmetries.}}
  	We first consider a family of 1D models to introduce the notion of modulated symmetries.
  	We consider stochastic cellular automaton dynamics, which allow for large-scale numerical simulations, although all of our models can be mapped to quantum models that realize the same set of symmetries.
  	We consider a chain of classical discrete spins that take values $s_x \in \{-S,\dots,S\}$ on each site $x=0,\ldots, L-1$. 
  	The dynamics is generated by local gates $G_x$, acting in a finite neighbourhood of site $x$. 
  	The effect of a gate of range $2\ell+1$ is described by a set of integers $n_i$ such that when applying $G_x$, the spins are updated as $s_{x+i} \to s_{x+i} \pm n_i$ with $i\in \{-\ell,\dots,\ell\}$.
  	We denote the gate as $G_x = \{n_i\}$.
  	The updates are applied probabilistically among those for which $|s_{x+i} \pm n_i| \leq S$, with symmetric transition rates
  	: at each application either (i) $s_{x+i} \to s_{x+i} + n_i$ is applied or (ii) its inverse, $s_{x+i} \to s_{x+i} - n_i$, or (iii) no update is made.
  	The full evolution is given by a random sequence of these gates~\cite{SP}.

  	We consider a family of models labeled by integers $q\geq1,p\geq 0$, defined by the gates
  	\begin{equation} \label{eq:H1d}
  		G^{(p,q)}_{x} = \left\lbrace n_{-1}, n_{0}, n_{+1} \right\rbrace = \left\lbrace q, -p, q \right\rbrace,
  	\end{equation}
  	acting on a three-site block centered around $x$
    (we can also use these to construct longer-range gates sharing the same symmetries~\cite{SP}).
  	For example $(p,q)=(2,1)$,  corresponds to a charge- and dipole-conserving spin chain studied in Refs. \onlinecite{Sala_2020,Khemani_2020,Rakovszky_2020,Moudgalya_2021}. 
  	For $2q\neq p$, on the other hand, these models do not conserve the total charge $\mathcal{Q}=\sum_j s_j$ or any of its higher moments.
    Nevertheless, there still exist some global conserved quantities, which we now construct~\footnote{In the following, we do not take into account the additional $\mathbb{Z}_2$ parity symmetries the following models can also have, as they do not affect the long-time behavior of spin correlations.}. 
    
    Consider the general ansatz $\mathcal{Q}_{\{\alpha_j\}}\equiv \sum_j \alpha_j s_j$. 
  	Then $\mathcal{Q}_{\{\alpha_j\}}$ is a conserved quantity for the evolution generated by $G^{(q,p)}$ if and only if $\{\alpha_j\}$ fulfills the recurrence relation
  	\begin{equation} \label{eq:rec}
  		\alpha_{j+2}=\frac{p}{q}\alpha_{j+1} - \alpha_{j}.
  	\end{equation}
  	%
%  	and thus for generic $q,p$, these symmetries correspond to unitary representations of the additive group $\mathbb{R}$ (similar to lattice translations), with integer spectrum only for $p/q\in \mathbb{Z}$.
  	%
%  	In this case, they lead to global U$(1)$ symmetries~\cite{Woit_17}.
  	%
%     of this difference, we will refer to all of them as U$(1)$ symmetries.  
  	%
  	As a linear recurrence, Eq.~\eqref{eq:rec}  equation admits the general solution
  	\begin{equation}\label{eq:rec_sol}
  	    \alpha_j = \frac{1}{r_1-r_2}\left[ (\alpha_1-\alpha_0r_2)r_1^j + (r_1\alpha_0-\alpha_1)r_2^{j}\right],
  	\end{equation}
  	where $r_1,r_2$ are the roots of the associated characteristic equation
  	\begin{equation} \label{eq:char}
  		r^2 - \frac{p}{q}r+1=0.
  	\end{equation}
  	The solutions are parameterized by the initial conditions $\alpha_0, \alpha_1$~\footnote{Equivalently, one can consider the boundary conditions $\alpha_0,\alpha_{L-1}$.}, which implies that the model Eq.~\eqref{eq:H1d} has at most two linearly-independent conserved quantities of this kind.  
  	Note that if $q$ divides $p$ then $\mathcal{Q}_{\{\alpha_j\}}$ has an integer spectrum and thus generates a representation of the group U$(1)$; otherwise the symmetry is a unitary representation of the additive group $\mathbb{R}$~\footnote{In general, one could enforce an integer spectrum via the system size dependent normalization $\mathcal{Q}_{\{\alpha_j\}}\to q^L\mathcal{Q}_{\{\alpha_j\}}$. }.
  	%
  	%In the following, we generically refer to them as U$(1)$ symmetries.
  	%
  	As the second order polynomial in Eq.~\eqref{eq:char} is palindromic or self-reciprocal~\cite{Roman_95}, its two roots $r_1,r_2$ are inverses of each other, $r_2=1/r_1$.  
  	Thus, three different scenarios can occur, depending on the ratio $p/q$:
  	% \pab{I think this classification might be exhaustive for physical systems. Think about it. $\log(x)$ or $\sqrt{x}$ ? or any other fundamental function missing?}
  	%
  	\paragraph{(i) Dipole conservation.} If $2q = p$ then $r_2=r_1=1$, which leads to a general solution of the form $\alpha_j=a_0+a_1j$; this reproduces the conservation of charge and dipole moment.
  	Although conserving higher moments require longer-range gates which implies characteristic polynomials of higher degree, one would again find that $r=1$ is the only root, so that $\alpha_j=\sum_{n=0}^m a_n j^n$. 
 
  	\paragraph{(ii) (Quasi-)periodic modulation.}
  	If $2q > p$, then Eq.~\eqref{eq:rec} has two complex solutions $\mathrm{e}^{\pm \mathrm{i} k^*}$ lying on the unit circle with $k^*=\arccos(\frac{p}{2q})$~\footnote{This can be generalized to longer-range gates using results from Ref.~\onlinecite{Konvalina2004PalindromePolynomialsWR}}. 
	A general solution of Eq.~\eqref{eq:rec} then takes the form $\alpha_j = a\mathrm{e}^{\mathrm{i}k^*j} + b\mathrm{e}^{-\mathrm{i}k^*j} = A\cos(k^* j + \phi)$ with constants $a,b$ (equivalently, $A,\phi$) fixed by $\alpha_0, \alpha_1$. 
  	Thus, while the total charge is not conserved, some finite momentum component of it is. 
  	However, while the recursion relation can always be solved in a system with open boundary conditions (OBC), the corresponding momentum mode might not exist in a finite system with periodic boundaries (PBC). 
  	Indeed, we could search directly for a conserved quantity of the form ${\mathscr{s}}_{{k}} \equiv \sum_j \mathrm{e}^{\mathrm{i}kj} s_j$, by plugging the ansatz $\alpha_j = \mathrm{e}^{\mathrm{i}kj}$ into Eq.~\eqref{eq:rec}, which then becomes $\chi(k)	\equiv \cos(k)-\frac{p}{2q}=0$. 
  	We can distinguish two possibilities, depending on whether the solution $k=k^*$ is a rational multiple of $\pi$ or not. 
  	According to Niven's theorem~\cite{niven_1956}, the former is the case if and only if $\frac{p}{2q}\in \{0,\pm \frac{1}{2},\pm 1\}$; in this case, the modulation $k^*$ is commensurate, having a finite periodicity on the lattice, and the symmetry is exact for some finite system sizes that are integer multiples of its period. 
  	In the more general case, however, $k^*$ is incommensurate, the modulation is {\it quasi-periodic}, and the conserved quantity does not exist for any finite system with PBC. 
  	Nevertheless, for sufficiently large systems, there will be momentum modes that are almost conserved, and the symmetry re-emerges in the thermodynamic limit.

  	\paragraph{(iii) Exponentially localized.}
  	For $2q < p$, the solutions $r_{1,2}$ are real, positive and non-degenerate.
  	This implies $r_1 > 1$ and $r_2 = 1/r_1 < 1$ (one could think of these as imaginary momenta, $r = \mathrm{e}^{\pm k}$). 
  	Plugging this into Eq.~\eqref{eq:rec_sol}, gives one solution that is exponentially decreasing with $j$ and another that is exponentially increasing. 
  	In this case, it is more appropriate to instead label solutions by the two endpoints, $\alpha_0$ and $\alpha_{L-1}$~\cite{SP}, rather than $\alpha_0$ and $\alpha_1$.
  	We can then think of the two conserved quantities as being exponentially localized at the two boundaries. 
  	Note that in this case, it is not possible to satisfy the recursion relation with PBC. 

  \textit{\textbf{Hydrodynamic description.}}
 Continuous symmetries provide long-lived modes that dominate the dynamics at long times; this idea is at the base of hydrodynamics~\cite{forster2018hydrodynamic,chaikin_lubensky_1995,10.21468/SciPostPhysLectNotes.18}. 
 %
% How does hydrodynamics change in the presence of unconventional modulated symmetries?
 %
% The answer was investigated recently 
Recent works investigated how hydrodynamics changes in the presence of multipole and subsystem symmetries~\cite{Feldmeier_2020,Gromov_2020,Morningstar_2020,Iaconis_2019,Iaconis_2021,Zhang_2020}, leading to sub-diffusive transport---with additional logarithmic corrections in the latter case.
Our goal now is to generalize these results to the exotic modulated symmetries discussed above. 

%We now generalize these results to the exotic modulated symmetries discussed above. 
 %
% The main quantity of interest are the ``infinite temperature'' spin-spin correlations $C(\vec{r},t) \equiv \overline{s_{\vec{r}}(t) s_{\vec{0}}(0)}/C_{\textrm{th}}$, where $\overline{(\dots)}$ denotes averaging over all randomly chosen initial spin configurations and circuit realizations, and $C_{\textrm{th}}=S(S+1)/3$. 
 
  The main quantity of interest are the ``infinite temperature'' spin-spin correlations $C(\vec{r},t) \equiv \overline{s_{\vec{r}}(t) s_{\vec{0}}(0)}$, where $\overline{(\dots)}$ denotes averaging over all randomly chosen initial spin configurations and circuit realizations. 
 In a system with dynamical exponent $z$ (ignoring, e.g., logarithmic corrections) this quantity has a scaling form at long times and large distances, $C(\vec{r},t) = t^{-1/z} f(\vec{r} / t^{1/z})$.
  	
  	Consider now periodically modulated symmetries corresponding to conserved momentum components of the total spin.
    These can be identified by the vanishing of some characteristic function, $\chi(\vec{k}) = 0$. 
  	To understand the dynamical consequences, we will assume a description in the spirit of linear hydrodynamics, which provides a closed linear equation of motion for some sufficiently coarse-grained version of the spin density. 
  	In momentum space, this can be written as $\partial_t {\mathscr{s}}_{\vec{k}}(t) = - \omega(\vec{k}) {\mathscr{s}}_{\vec{k}}(t)$. 
  	The key difference from more usual hydrodynamic descriptions is that we cannot simply expand the ``imaginary frequency'' $\omega(\vec{k})$ near $\vec{k} \approx \vec{0}$.
  	Instead, we have to take into account the slow modes at finite momenta originating from the modulated symmetries. 
    The influence of finite (lattice-scale) momentum components in the BZ on long-time / large-distance correlations can be seen as a manifestation of UV/IR-mixing in these models~\cite{Seiberg_2020,gorantla2021lowenergy}.
  	
  	To obtain $\omega(\vec{k})$, we require that: (i) $\omega(\vec{k}) \geq 0$, (ii) $\omega(\vec{k}) = 0\Leftrightarrow\chi(\vec{k}) = 0$ %such that ${\mathscr{s}}_{\vec{k}}(t)$ remains constant,
  	and (iii) $\omega(\vec{k})$ is analytic around these points. 
   	A natural approximation that satisfies all these requirements and should correctly capture the leading order behavior in the regimes where $\omega(\vec{k}) \approx 0$ is $\omega(\vec{k}) \sim |\chi(\vec{k})|^2$. One can check that this approximation correctly captures the known behavior in a variety of models, including those with subsystem symmetries~\cite{SP}. 
  	
  	Within linear response, the spin-spin correlator should behave as the Green's function of this equation of motion~\cite{forster2018hydrodynamic}, $C(\vec{r},t) = \int \mathrm{d}^dk\!\mathop{}\mathrm{e}^{\mathrm{i}\vec{k}\cdot\vec{r} - \omega(\vec{k}) t}$. For the autocorrelation, $\vec{r} = \vec{0}$, we can rewrite this as 
  	\begin{equation}\label{eq:C_from_DOS}
  	    C(\vec{0},t) = \int_0^\infty \mathrm{d}\omega\!\mathop{}\rho(\omega) \mathrm{e}^{-\omega t}.
  	\end{equation}
  	The long-time decay is therefore determined by the density of states (DOS), $\rho(\omega)$, near $\omega \approx 0$~\cite{Vijay2021}. 
  	
\begin{figure}[t!]
  		\centering
  		\includegraphics[width=\linewidth]{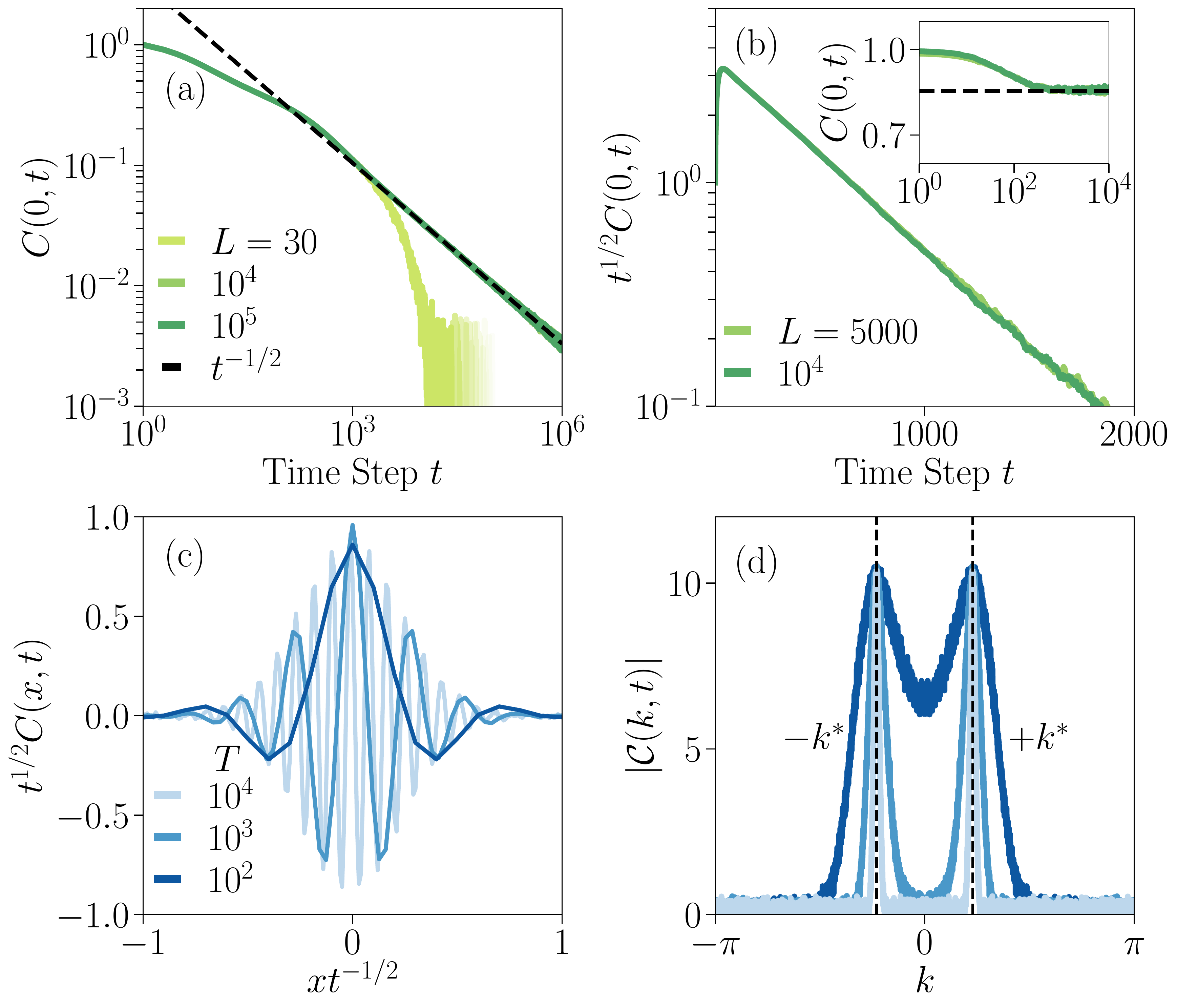}
  		
  		\caption{\textbf{Long-time behavior in 1D.} Evolution of the spin-spin correlator for the 1D models in Eq.~\eqref{eq:H1d}. (a) Evolution of the autocorrelator $C(0,t)$ for quasi-periodic symmetries: data for $S=5$ and $(p,q)=(3,2)$; (b) exponentially-localized symmetries with $S=10$ and $(p,q)=(3,1)$. The inset shows the boundary correlation which is lower bounded by Mazur's bound (black dashed line). (c-d) Spatial correlations for the model in panel (a): ``Dressed'' scaling collapse of $C(x,t)$ (c) and its spatial Fourier transform $\mathcal{C}(k,t)$, which becomes increasingly peaked at $k= \pm k^*$ (d).}
  		\label{fig:ffig1}
  	\end{figure}

  	Consider $G^{(p,q)}$ with $2q>p$. 
  	Near the conserved momentum $k^*$ we have $\omega(k^*+\delta k) \sim \delta k^2$.  %$|\chi(k)|^2 = |\cos(k^* + \delta k) - \frac{p}{2q}|^2 \sim \delta k^2 $.
  	This gives rise to a DOS $\rho(\omega) \sim \omega^{-1/2}$; inserting this into Eq.~\eqref{eq:C_from_DOS} yields a diffusive scaling, $C(0,t) \sim t^{-1/2}$. 
  %	a diffusive scaling despite the fact that the total spin is not conserved in these systems; 
  	This is consistent with our numerical results, shown in Fig.~\ref{fig:ffig1}a for $(p,q) = (3,2)$, and for larger gates~\cite{SP}.
  	Nevertheless, as this is not an exact symmetry for PBC, $C(0,t)$ is expected to decay exponentially at sufficiently long times (see data for $L=30$ in Fig.~\ref{fig:ffig1}a).
  	The situation changes in the dipole-conserving limit $2q=p$. 
  	In this case, $k^* = 0$, so the leading contribution vanishes and we instead have $\omega(k) \sim k^4$ near $k\approx0$. 
  	This gives $\rho(\omega) \sim \omega^{-3/4}$ and $C(0,t) \sim t^{-1/4}$, recovering the known sub-diffusive scaling~\cite{Guardado_Sanchez_2020,Feldmeier_2020,Morningstar_2020,Gromov_2020,Zhang_2020}. 
  	
  	The role of finite-momenta modes becomes much more apparent when we consider the spatial structure of the correlations. 
  	Taking into account the slow modes around $k \approx \pm k^*$, we obtain $C(x,t) \sim t^{-1/2} \mathcal{N}(x/\sqrt{t}) \cos(k^* x)$, i.e., diffusive behavior modulated by a factor that oscillates at the microscopic scale $1/k^*$, resembling the ``UV/IR'' phenomenology~\cite{Seiberg_2020,gorantla2021lowenergy,you2020fracton}.
  	This behavior is numerically verified in Fig.~\ref{fig:ffig1}(c,d).%, with the short wavelength oscillations in $C(x,t)$ appearing as a non-vanishing contribution in its spatial Fourier transform $\mathcal{C}(k,t)$ (see Fig.~\ref{fig:ffig1}.(d)). 

  	We can also apply our approximation to models with exponentially localized symmetries. 
  	In this case, $\omega(k) \sim |\chi(k)|^2$ is finite everywhere, which indicates an exponential decay of correlations.
  	Nevertheless, there can be a correction coming from the large density of states near the minimum of $\omega(k)$. 
  	To see this, consider again the model Eq.~\eqref{eq:H1d}, but this time with $2q < p$. 
  	The dispersion has a minimum at $k=0$ and expanding around it we find $\omega(k) \approx (\frac{1}{2}k^2-k_{0}^2 )^2$, with $k_{0}^2 \equiv \frac{2q-p}{2q}$. 
  	Integrating over $k$ we find an analytical solution, which has the long-time asymptotic form $C(0,t) \sim \mathrm{e}^{-k_{0}^4 t}/\sqrt{t}$.
  	In Fig.\ref{fig:ffig1}b, we numerically verify this for correlations in the bulk of the system.
  	However, the exponentially localized symmetries have a strong effect on the dynamics near the boundary, leading to infinitely long-lived correlations,
  	%
%  	These are lower bounded by $(1-1/r^2)S(S+1)/3$ in the thermodynamic limit, 
    as one can prove using Mazur's inequality~\cite{SP,MAZUR1969533,SUZUKI1971277} (dashed line in the inset of Fig.\ref{fig:ffig1}b). 

	\textit{\textbf{Generalization to higher-dimensions.}}
	We now generalize our discussion to 2D systems.
	We begin by constructing a model which features the quasi-periodically modulated symmetries discussed above. 
	However, in this case, the conserved momentum components will not only lie at isolated points in the Brillouin zone (BZ), but extend along continuous lines.
	%, reminiscent of a `Fermi surface' in our classical model. 

\begin{figure}[t!]
  	\centering
  	 \makebox[\linewidth][c]{\includegraphics[width=0.5\textwidth]{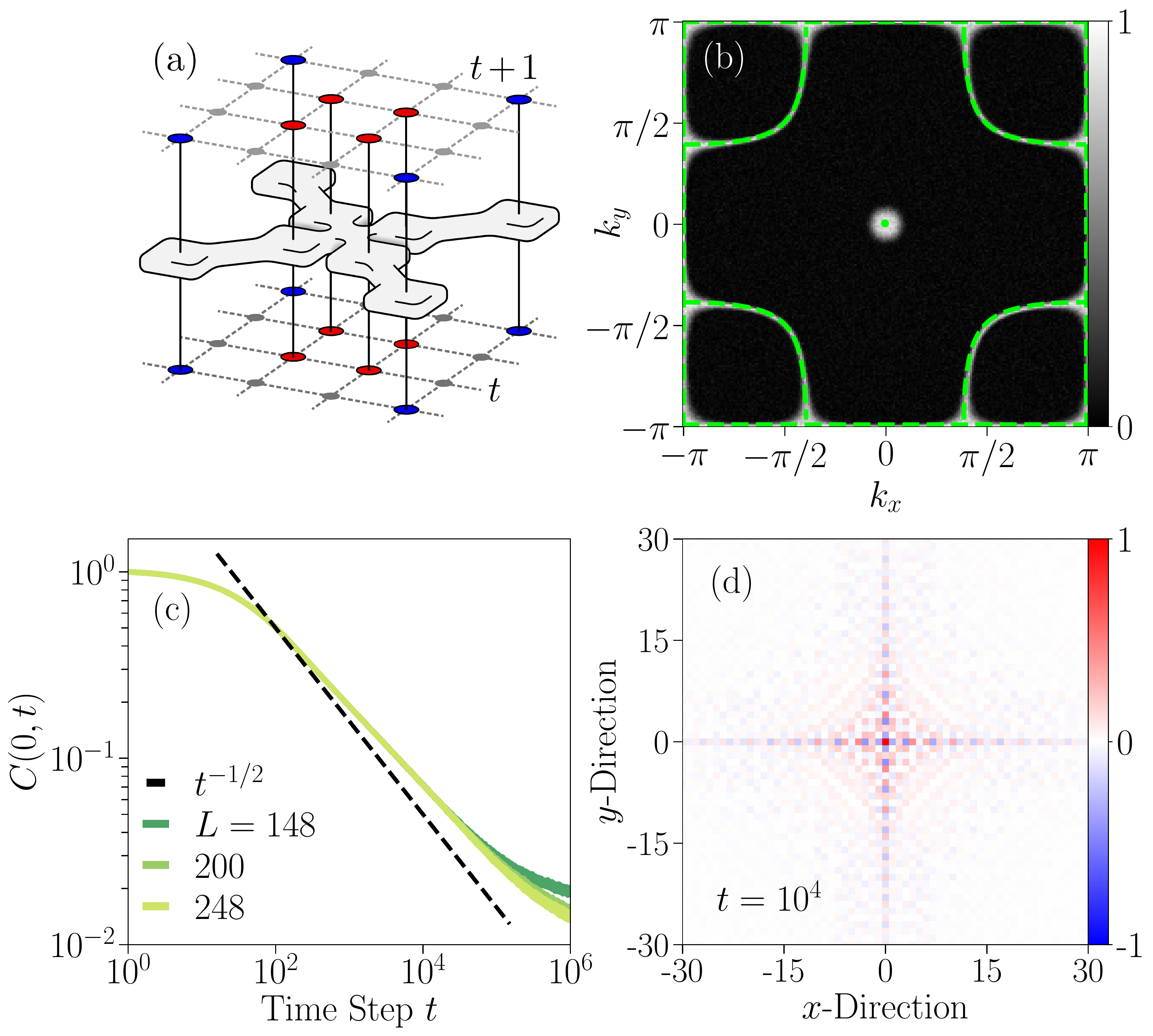}}
  	\caption{\textbf{Long-time dynamics in 2D.} (a) Schematics of a local gate  corresponding to Eq.~\eqref{eq:Hex}. (b) Two-dimensional correlation function $\mathcal{C}(\vec{k},t)$ within the Brilouin zone normalized to the interval $[0,1]$. The solution of Eq.~\eqref{eq:2dchi} is shown in a green (dashed) line. (c) Autocorrelation decay $C(0,t)$ for different linear system sizes $L$. (d) Spatial charge correlation $C(\vec{x},t)$ shown for $L=200$ and $t=10^4$ normalized to the interval $[-1,1]$. %A highly fluctuating local structure concentrating around the two coordinate axes is clearly visible and remains for longer times. 
  	Correlations are concentrated along the two axes and show oscillations on lattice scales that survive for long times (Data shown for $S=3$).}
  	\label{fig:ffig2}
\end{figure}

  	In our microscopic model, local gates $G_{x,y}$ act on a $4\times 4$ block of a 2D square lattice in the vicinity of the site with coordinates $x,y$. 
  	The gate is again specified by a set of integers, such that $G_{x,y}: s_{x+i,y+j} \to s_{x+i,y+j} \pm  n_{i,j}$, with
  	\begin{align}\label{eq:Hex}
  	    G &= \{n_{0,0},n_{0,3},n_{3,0},n_{3,3},n_{1,1},n_{1,2},n_{2,1},n_{2,2}\} = \nonumber \\ 
  	    &= \{1,1,1,1,-1,-1,-1,-1\},
  	\end{align}
  	i.\,e., it symmetrically moves four charges between the central $2\times2$ plaquette and the four outer corners as illustrated in Fig.~\ref{fig:ffig2}a. 
  	
    This model has many U$(1)$ symmetries:
    It conserves both the total charge, $\mathcal{Q}^{(0)}=\sum_{\vec{r}}s_{\vec{r}}$, its dipole moment $\bm{\mathcal{Q}}^{(1)}=\sum_{\vec{r}}\vec{r}s_{\vec{r}}$ and the $\mathcal{Q}^{(2)}_{xy}, \mathcal{Q}^{(2)}_{x^2-y^2}$ components of the quadratic moment (however, it does not conserve $\mathcal{Q}^{(2)}_{x^2+y^2}$).
    Moreover, it conserves the {\it staggered magnetization} along all rows and columns: $\mathcal{S}_{x_0}=\sum_{y}(-1)^ys_{x_0,y},\,\,\mathcal{S}_{y_0}=\sum_{x}(-1)^xs_{x,y_0}$.
  	%For example, a one-dimensional system with gates $G_{x}=\{n_0,n_1\}=\{+1,+1\}$ acting on two consecutive sites, would have such conservation law.
  	%
  	However, these do not exhaust the set of modulated symmetries of the model.
    To detect additional modulated conserved quantities $\mathcal{Q}_{\{\alpha_{\vec{r}}\}}=\sum_{\vec{r}}\alpha_{\vec{r}}s_{\vec{r}}$, we can follow the steps of the previous section and look for non-trivial solutions of the associated two-dimensional recurrence relation
  	\begin{align} \label{eq:rec2D2_main}
  	\sum_{i,j} n_{i,j} \alpha_{x+i,y+j}=0.
  	\end{align}
   Although a complete analytical solution of this equation might be feasible (e.g., via generating functions~\cite{10.2307/2984940} or rewriting it as a Sylvester equation~\cite{10.2307/3215145}), the procedure can be quite involved.
    Instead we use the ansatz $\alpha_{\vec{r}}=\mathrm{e}^{\mathrm{i} \vec{k}\cdot \vec{r}}$ and focus on periodically modulated symmetries (one can check that this model has no exponentially localized ones of the form $\alpha_{\vec{r}}=\mathrm{e}^{\vec{k}\cdot \vec{r}}$).
    Eq.~\eqref{eq:rec2D2_main} then reduces to
   \begin{align} \label{eq:2dchi}
   \nonumber
       \chi(\vec{k})&\equiv \sum_{a,b} n_{a,b} \mathrm{e}^{\mathrm{i} (a k_x + b k_y)} \propto  8\cos\left(k_x/2\right)\cos\left(k_y/2\right) \times \\ \times &\big[\cos(k_x) + \cos(k_y)
       -2\cos(k_x)\cos(k_y)\big] = 0.
       %\mathrm{e}^{\frac{i}{2}(k_x+k_y)} = 0.
   \end{align}

The solutions of $\chi(\vec{k})=0$ are highlighted in green in Fig.~\ref{fig:ffig2}b.
The conservation of total charge corresponds to the mode ${\mathscr{s}}_{\vec{0}}$ at $\vec{k}=(0,0)$, while the staggered subsystem symmetries $\mathcal{S}_{x},\mathcal{S}_{y}$ show up as the lines $k_y=\pi$ and $k_x=\pi$ respectively. 
Moreover, we find a set of contour lines (forming a closed loop in the Brillouin zone) along which the second line of Eq.~\eqref{eq:2dchi} vanishes.
As in 1D, each of these corresponds to an exact symmetry for OBC, whose total number scales with the linear system size $\mathcal{O}(L)$~\cite{SP}.
However, most points along these lines are not realized exactly in a finite system with PBC, but they become only exact symmetries in the thermodynamic limit, leading to an (infinite dimensional) emergent symmetry group.
While straight lines in the BZ correspond, upon inverse Fourier transformation, to symmetry operators that act along columns or rows on the lattice, linear combinations of symmetries lying along these contours do not seem to lead to subsystem symmetries even in the thermodynamic limit~\cite{SP}.

As we saw, the asymptotic decay of $C(\vec{0},t)$ is governed by the DOS near $\omega \approx 0$.
For the 2D model in Eq.~\eqref{eq:Hex}, $\rho(\omega)$ picks up contributions from various parts of the BZ (see Fig.~\ref{fig:ffig2}b)~\cite{SP}. 
We find that the leading contributions arise from the five points where multiple lines of conserved momenta meet, $(k_x,k_y) = (\pi,\pi), (\pi,\pm k^*), (\pm k^*,\pi)$. 
Around all of these points,  $\rho(\omega) \sim \omega^{-1/2} \log(\omega)$ and consequently $C(\vec{0},t) \sim t^{-1/2} \log(t)$: sub-diffusion with a logarithmic correction, similarly to the case of $U(1)$ subsystem symmetries~\cite{Iaconis_2019} (the remaining parts of the BZ provide a subleading $\rho(\omega)\sim\omega^{-1/2}$ contribution). 
Indeed, in Fig.~\ref{fig:ffig2}c we observe numerically that correlations decay slower than $t^{-1/2}$. 
The finite momentum contributions also lead to spatial oscillations of $C(\vec{x},t)$  at short scales (see Fig.~\ref{fig:ffig2}d), which can be clearly identified in its Fourier transform $\mathcal{C}(\vec{k},t)$ shown in Fig.~\ref{fig:ffig2}b, concentrated along the solutions of $\chi(\vec{k})=0$. 
Another consequence is that $C(\vec{r},t)$ does not have full rotational invariance, but instead concentrates around the two coordinate axes.
%
%All this together implies that $C(\vec{x},t)$ does not have a simple ``dressed'' scaling form like the one we found in 1D, or generally expected for systems with only higher-moment conservation.

\begin{figure}[t!]
  		\centering
  		\makebox[\linewidth][c]{\includegraphics[width=\linewidth]{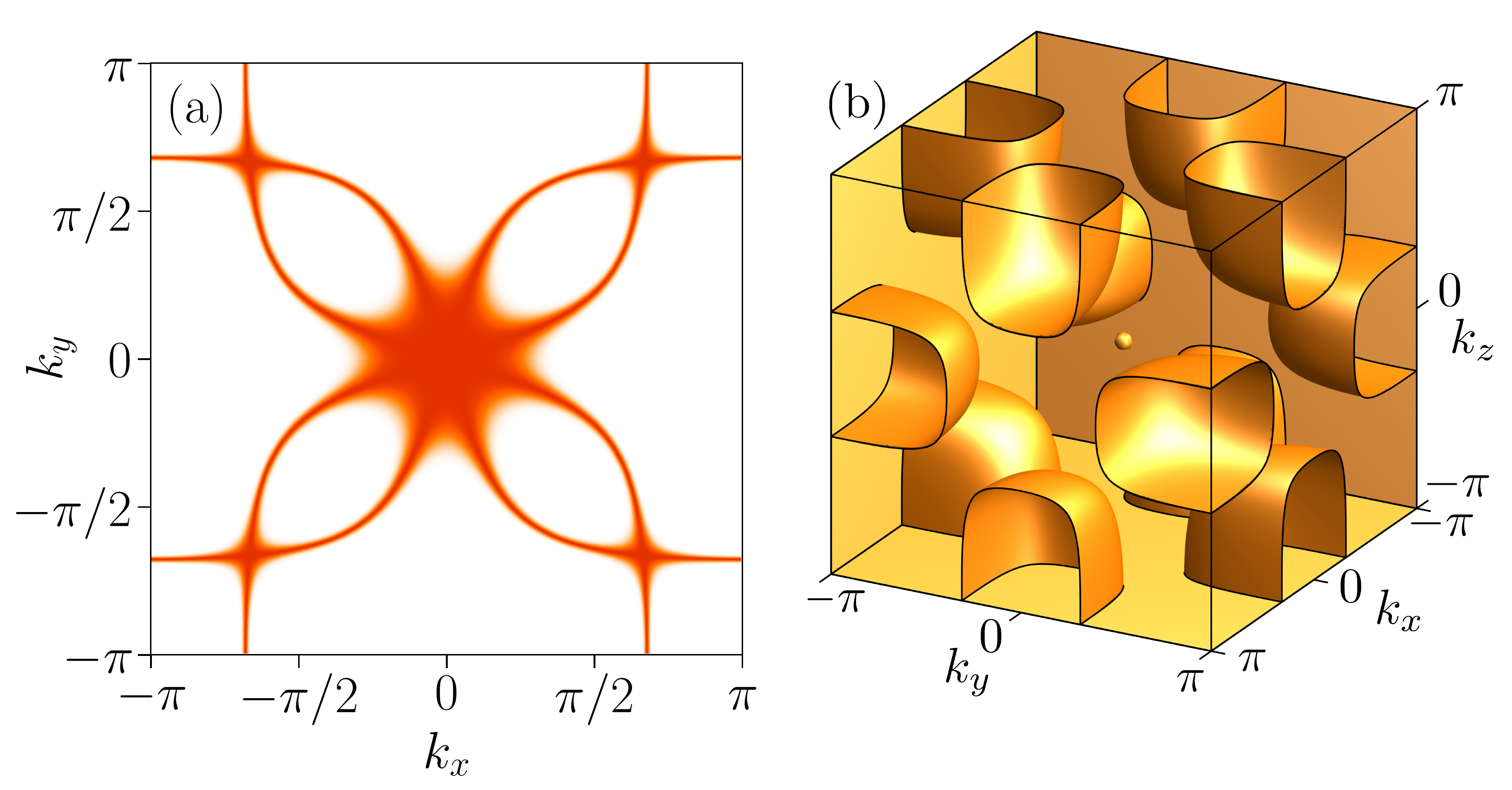}}
  		\caption{ \textbf{Higher-dimensional systems with (quasi)-periodic symmetries.} (a) Example of a 2D system with $5\times 5$ gates and $\omega(\vec{k}\approx \vec{0})\approx (k_x^4+k_y^4 -11k_x^2k_y^2)^2$. The figure shows $\mathrm{e}^{-|\chi(\vec{k})|^2}$. (b) 3D generalization of the 2D model in Eq.~\eqref{eq:Hex}. The figure shows the solutions of $\chi(\vec{k})=0$.}
  		\label{fig:ffig4}
\end{figure}

%Although we focused on a particular model, 
One can construct many other models which exhibit conserved momenta along various shapes in the BZ; 
%
%In fact, one could reverse engineer microscopic models from the knowledge of $\chi(k_x,k_y)$.
%
an example is shown in Fig.~\ref{fig:ffig4} for $5\times 5$ gates~\cite{SP}.
The construction can also be easily extended to higher dimensions.
E.g., in 3D, one can find exact conserved quantities lying in intersecting 2D manifolds in momentum space~\cite{SP} as we show in Fig.~\ref{fig:ffig4}b.
These symmetries lead to different scalings for the correlations, depending on the details of $|\chi(\vec{k})|$.
The decay is at least as slow as $C(\vec{0},t) \sim t^{-1/2}$, coming from the fact that expanding the dispersion along a co-dimension 1 hypersurface (a line in 2D or a surface in 3D) is formally similar to an expansion in one dimension. 
However, the actual behavior can be much slower than this. 
For example, in 2D, points where many conserved lines intersect (as in Fig.~\ref{fig:ffig4}a), or ones where lines touch (rather than cross) lead to strongly sub-diffusive dynamics~\cite{SP}.

A natural question is whether it is possible to derive these results directly from a continuum hydrodynamic formulation, similar to Ref.~\onlinecite{Gromov_2020}.
In 1D, we can expand around $k^*$, as given by the microscopic model, and construct a continuum theory for a scalar field $\phi$ that is invariant under the shift symmetry $\phi(x) \to \phi(x) + \alpha(x)$ with $\alpha(x)$ satisfying $(\partial_x^2 + (k^*)^2)\alpha(x)=0$, a continuum version of the recurrence relation in Eq.~\eqref{eq:rec}.
This resembles the $(1+1)$D UV-theories introduced in Ref.~\onlinecite{Lake_2021}.
The situation is much more complicated in higher dimensions, where there are infinitely many symmetries, encoding the intricate shape of conserved modes in the BZ. 
In this case, we would require invariance under any shift $\alpha(\vec{x})$ whose Fourier transform satisfies $\chi(\vec{k}) \tilde{\alpha}(\vec{k}) = 0$. 
The derivation of an appropriate hydrodynamic field theory in this case is an interesting challenge that we leave for future work.

Finally, one can also obtain models that exhibit exponentially localized (at the boundary of the 2D lattice) symmetries, whose total number scales with the linear system size~\cite{SP}.
An example of such systems is given by the $3\times 3$ gates $G= \{n_{0,0},n_{-1,0},n_{1,0},n_{0,1},n_{0,-1}\} = \{4,-1,-1,-1,-1\}$, for which an exact solution of the recurrence relation can be found in Ref.~\onlinecite{10.2307/3215145}.
This provides an example of a 2D model with $\mathcal{O}(L)$ exponentially localized symmetries at the boundary of the system; we leave the exploration of their effect on boundary dynamics to future work.
  
  	\textit{\textbf{Conclusions.}}
  	We introduced the notion of {\it spatially modulated symmetries} that generalize both multipole and subsystem symmetries.
  	We provided two new classes of such symmetries: quasi-periodically modulated and exponentially localized ones. 
  	The latter are relevant for the dynamics near the boundary, playing a role similar to strong zero modes. %, whose implications in 2D are worth exploring. 
  	The former lead to unusual behavior in bulk correlations: 
  	sub-diffusive decay and long-lived short-wavelength oscillations, provinding a hydrodynamic analog of the phenomenon of UV/IR mixing, making long-time dynamics sensitive to lattice-scale features. 
  	While here we discussed thermal correlations, these models also appear to host interesting examples of fragmentation which we plan to explore in a future publication.
  	
  	Although we focused on classical cellular automata, each of our models can be easily related to a corresponding quantum Hamiltonian,
  	by mapping a gate $G_{\vec{r}}$, characterized by integers $\{n_{\vec{a}}\}$, to a local Hamiltonian term $\bigotimes_{\vec{r}} \big( \hat{S}^{\operatorname{sgn}(n_{\vec{a}})}_{\vec{r}+\vec{a}} \big)^{|n_{\vec{a}}|}$. 
  	These quantum Hamiltonians possess the same set of symmetries as their classical counterparts; understanding their low-energy physics and its relationship to previous studies of UV/IR mixing, is an exciting challenge.
  	Moreover, 1D systems with such quasi-periodic symmetries might be realized as effective descriptions in the strong detuning limit of experimental realizations of the Aubry-André model~\cite{Schreiber_2015,Kohlert_2019}.
  	Finally, generalizing our analysis of long-time dynamics to 
  	  %models with U$(1)$ fractal symmetries~\cite{myersonjain2021pascals} and
  	  3D is another interesting open question.

  	\textit{\textbf{Acknowledgements.}}
    We thank Luca Delacretaz, Johannes Feldmeier, Adrian Franco Rubio and Sebastian Scherg for helpful discussions. We are grateful to Sagar Vijay for an insightful discussion regarding the decay of autocorrelations, and to Yizhi You for fruitful discussions and her comments on the manuscript. TR thanks Chaitanya Murthy for sharing with him some of his vast knowledge of Fourier transforms, among other topics.
     T.R. is supported by the Stanford Q-Farm Bloch Postdoctoral Fellowship in Quantum Science and Engineering. 
    TR acknowledges the hospitality of the Aspen Center for Physics, supported by National Science Foundation grant PHY-1607611 and the Kavli Institute for Physics, supported by the National Science Foundation under Grant No. NSF PHY-1748958. 
    F.P. acknowledges support from the European Research Council (ERC) under the European Union’s Horizon 2020 research and innovation programme (grant agreement No. 771537) and the Deutsche Forschungsgemeinschaft (DFG, German Research Foundation) under Germany’s Excellence Strategy EXC-2111-390814868 and TRR 80.\\
    
    \textit{Note added.} Recently, we became aware of related work by Oliver Hart, Andrew Lucas and Rahul Nandkishore~\cite{Toappear}.

  	%\bibliographystyle{plain}
%  	\bibliography{biblio}
  	\bibliography{modsym.bib}

  	\newpage
%  	\leavevmode \newpage
  	
\section{Appendix}
     
\section{Further examples of systems with periodically modulated symmetries}\label{app:models}

In this appendix we show some additional examples of 2D and 3D models with quasi-periodic symmetries.
The 2D models consist on size-$5$ gates $G_{\vec{r}}=\{n_{i,j}\}$ with $i,j\in \{-2,\dots,+2\}$, acting on a neighbourhood of a site $\vec{r}$.
$n_{i,j}$ are identified with the entries of a $5\times 5$ matrix.
The first set of models, shown in Table.~\ref{tab:Table1}, conserve all components of the quadrupole moment, including $\mathcal{Q}_{x^2+y^2}$, and are constructed to be $C_4$ symmetric.
Imposing these symmetries leads to the general matrix expression
\begin{equation}
   n_{i,j}= \begin{pmatrix*}[c]
    a&b&c&f&a\\
    f&g&h&g&b\\
    c&h&m&h&c\\
    b&g&h&g&f\\
    a&f&c&b&a
    \end{pmatrix*}.
\end{equation}
Different choices of the parameters $a,\ldots,m$ will in general correspond to different quasi-periodic modulated symmetries. 
In terms of these parameters, $\chi(\vec{k})$ takes the form
\begin{align} \label{eq:eqc4}
    \nonumber &\chi(\Vec{k})=4a\cos(2k_x)\cos(2k_y)+4b\bigl(\cos(k_x)\cos(2k_y)\\
    \nonumber &+\cos(2k_x)\cos(k_y)\big)+2c\bigl(\cos(2k_x)+\cos(2k_y)\big)\\
    &+4g\cos(k_x)\cos(k_y)+2h\bigl(\cos(k_x)+\cos(k_y)\big)+m.
\end{align}

\begin{table}[h!]
    \centering
    \begin{tabular}{|c|c|}\hline
    \multicolumn{2}{|c|}{Characteristic $\chi(\vec{k})$ \vphantom{$\sum_f^f$}}\\\hline
        \includegraphics[width=0.5\linewidth]{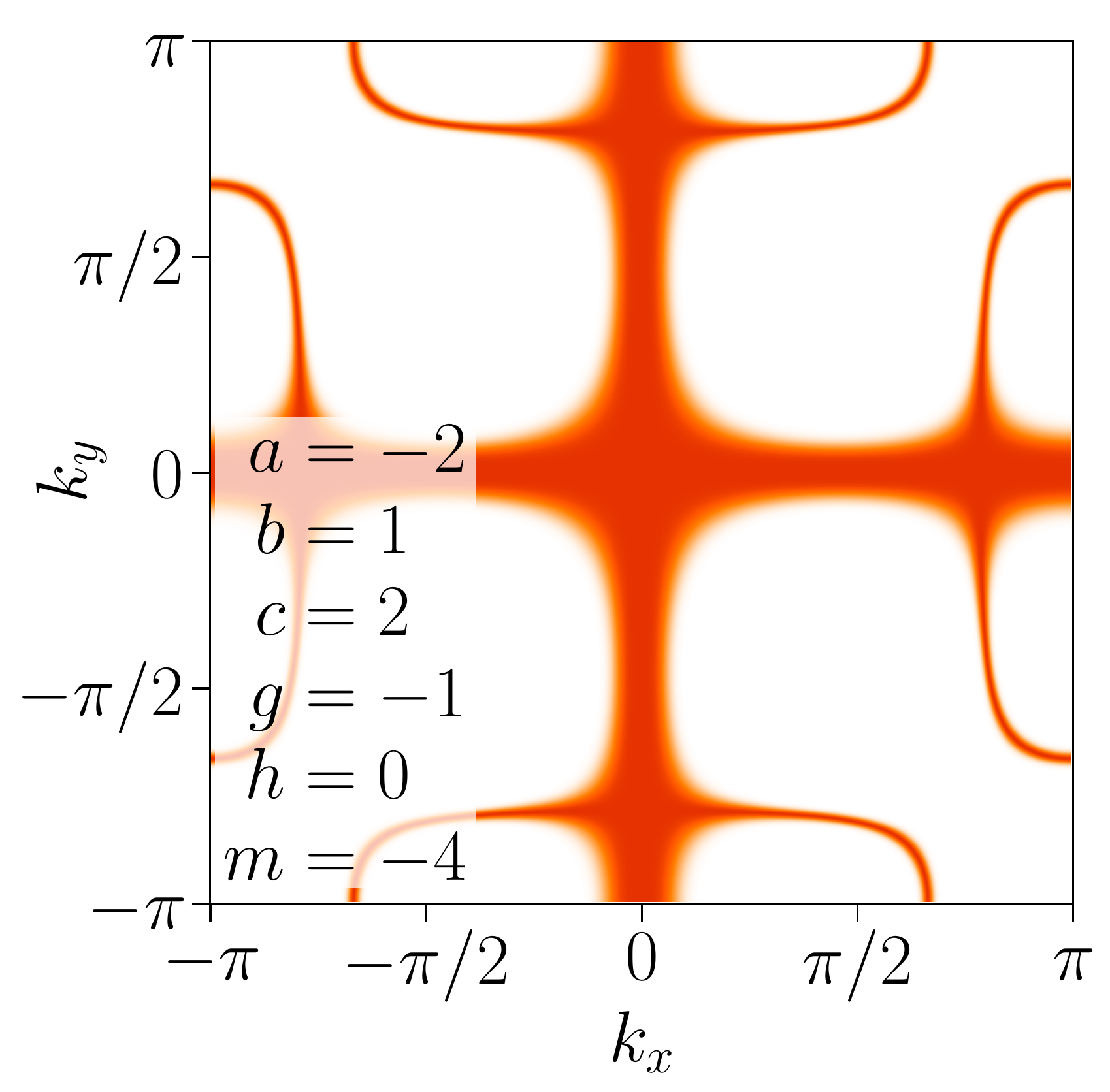}  &
        \includegraphics[width=0.5\linewidth]{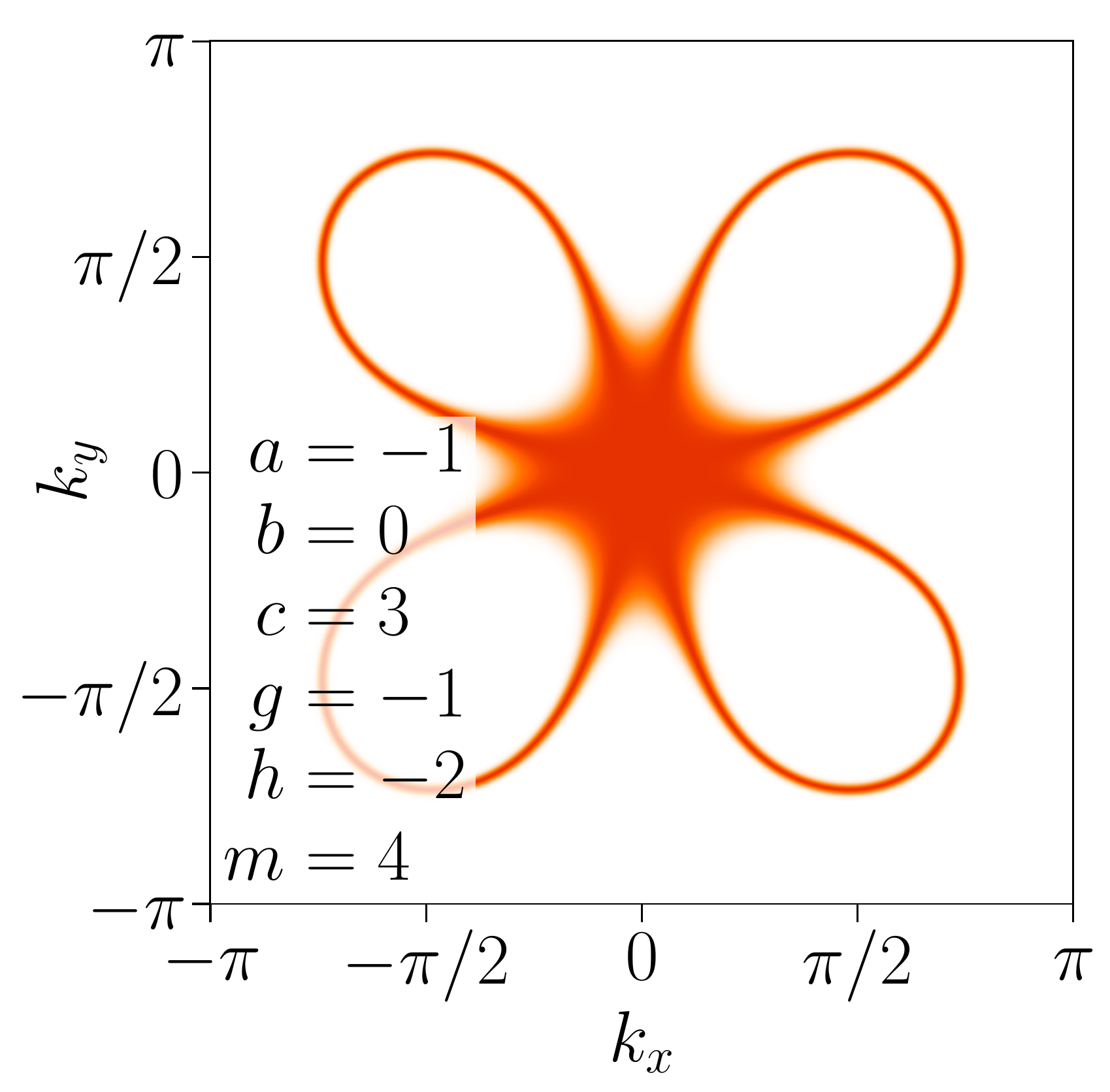}\\\hline
    \multicolumn{2}{|c|}{Leading order of $\omega(\vec{k}\approx\vec{0})$\vphantom{$\sum_f^f$}}\\\hline
        $(25k_x^2k_y^2)^2$ \vphantom{$\sum_f^f$} & $(k_x^4+k_y^4-17k_x^2k_y^2)^2$\\\hline
    \end{tabular}
    \caption{Examples of 2D systems with quasi-periodic symmetries and $C_4$ symmetric gates. The plot shows $\mathrm{e}^{-|\chi(\vec{k})|^2}$, with the analytical expression for $\chi(\vec{k})$ given in Eq.~\eqref{eq:eqc4}. The corresponding parameters are specified in the panel.}
    \label{tab:Table1}
\end{table}

Dropping the requirement of $C_4$ symmetry (and instead considering skew-centrosymmetric gates), while still preserving all quadratic moments leads to the general matrix expression

\begin{equation}
    n_{i,j}=\begin{pmatrix*}[r]
    a&b&c&-f&-a\\
    f&g&h&-g&-b\\
    c&h&0&-h&-c\\
    b&g&-h&-g&-f\\
    a&f&-c&-b&-a
    \end{pmatrix*},
\end{equation}
with 
\begin{align} \label{eq:eqnoc4}
    \nonumber &\chi(\Vec{k})=2 a  \sin(2 k_x) \cos(2 k_y)+ b\bigl(\sin(k_x - 2 k_y) \\ \nonumber&+ \sin(2 k_x + k_y)\big) + c\bigl(\sin(2 k_x) - \sin(2 k_y)\big) \\ \nonumber& +f\bigl(\sin(2 k_x - k_y) + \sin(k_x + 2 k_y)\big) \\&+2 g \sin(k_x)\cos(k_y)  + h\bigl(\sin(k_x) - \sin(k_y)\big).
\end{align}
Results for models of this type are shown in Table~ \ref{tab:Table2}.

\begin{table}
    \centering
    \begin{tabular}{|c|c|}\hline
    \multicolumn{2}{|c|}{Characteristic $\chi(\vec{k})$\vphantom{$\sum_f^f$}}\\\hline
        \includegraphics[width=0.5\linewidth]{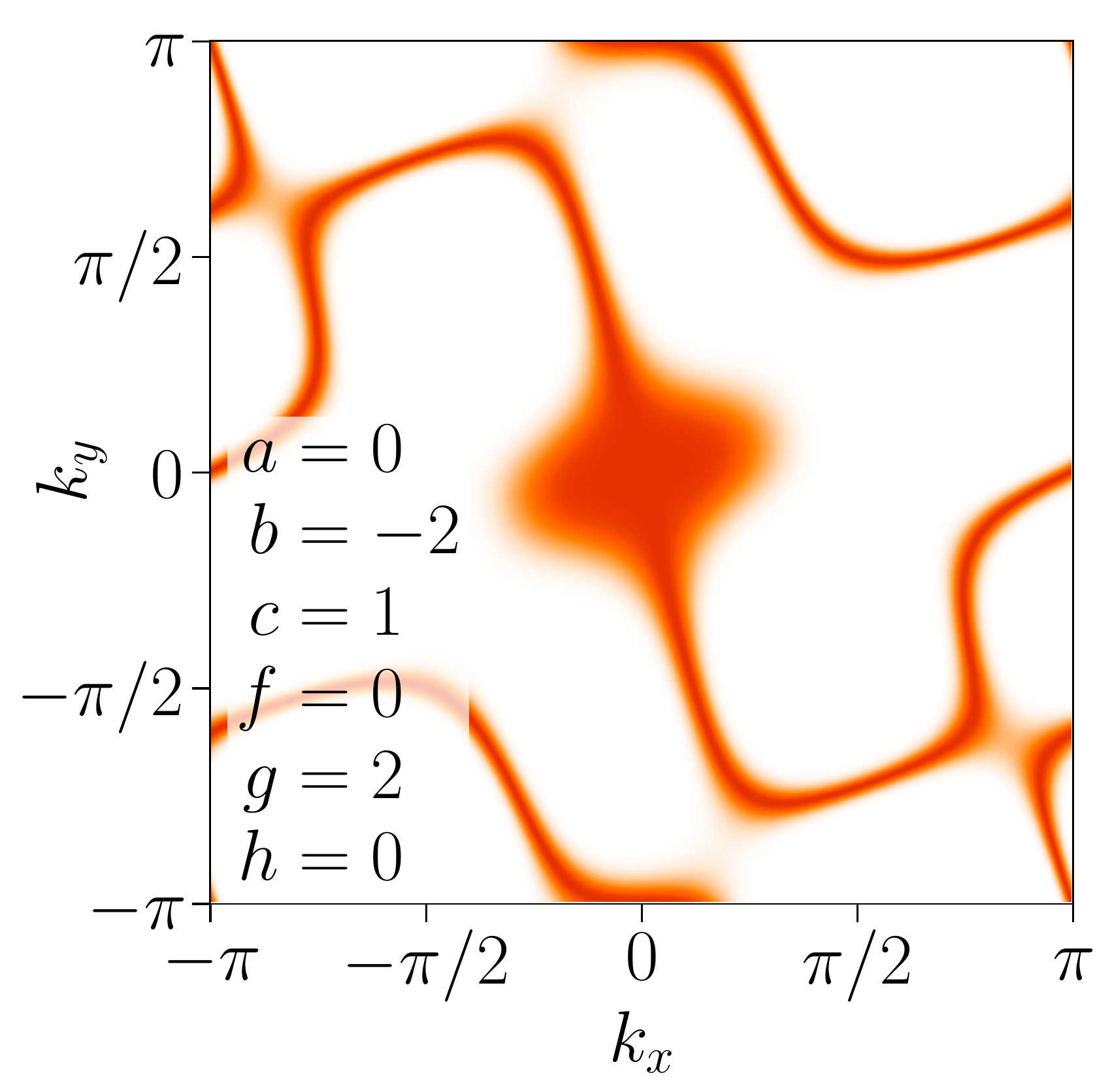}&
        \includegraphics[width=0.5\linewidth]{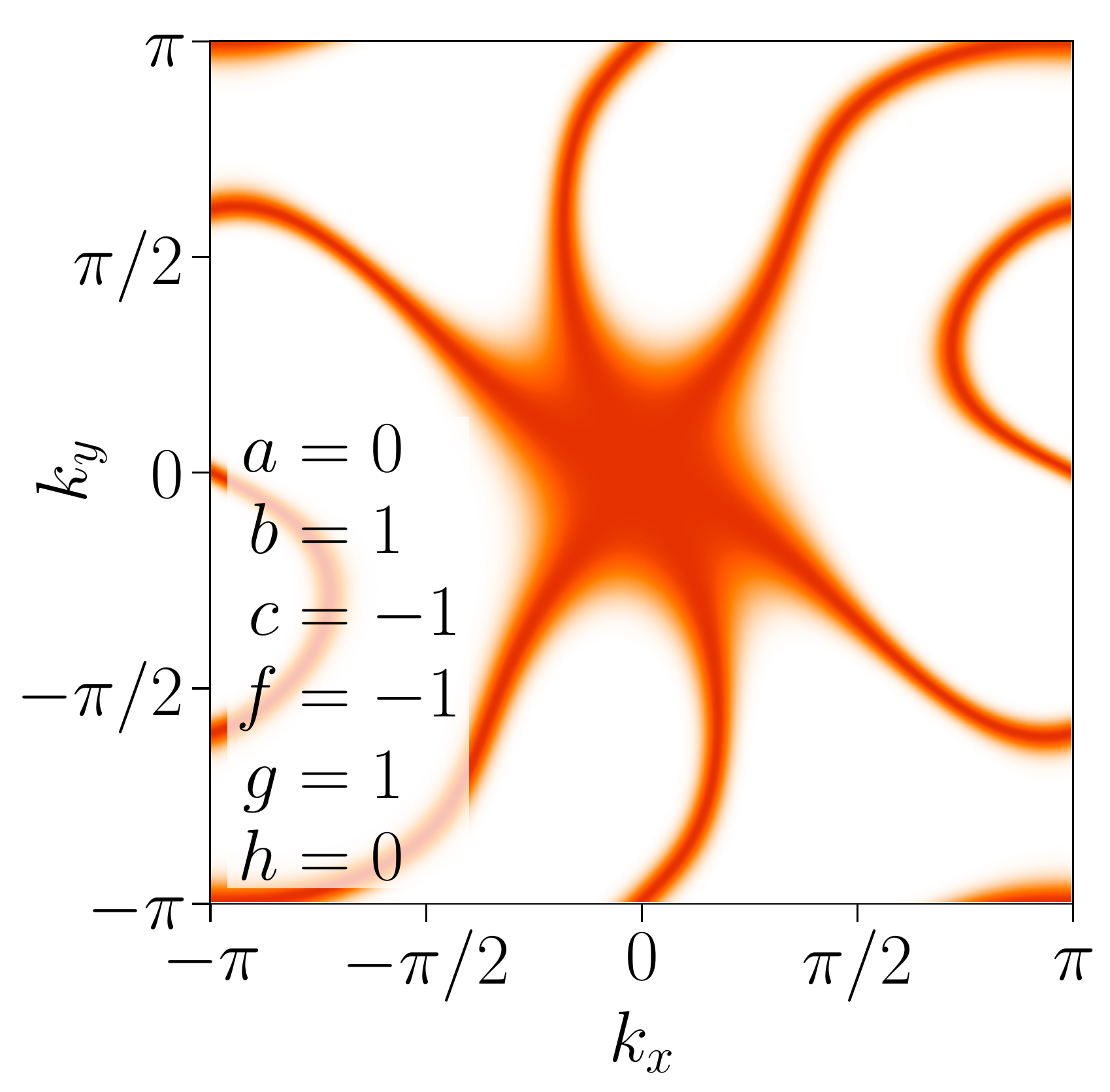}\\\hline 
    \multicolumn{2}{|c|}{Leading order of $\omega(\vec{k}\approx\vec{0})$\vphantom{$\sum_f^f$}}\\\hline
        $(k_x^3 + 2 k_x^2 k_y + 4 k_x k_y^2 - k_y^3)^2$ \vphantom{$\sum_f^f$}&
        $(k_x^3 - 2 k_x^2 k_y - k_x k_y^2 + k_y^3)^2$ \\\hline
    \end{tabular}
    \caption{Examples of 2D systems with quasi-periodic symmetries. The plot shows $\mathrm{e}^{-|\chi(\vec{k})|^2}$, with analytical expression  with the analytical expression for $\chi(\vec{k})$ given in Eq.~\eqref{eq:eqnoc4}. The corresponding parameters are specified in the panel.}
    \label{tab:Table2}
\end{table}

\begin{figure}
    \centering
    \includegraphics[width=0.35\linewidth]{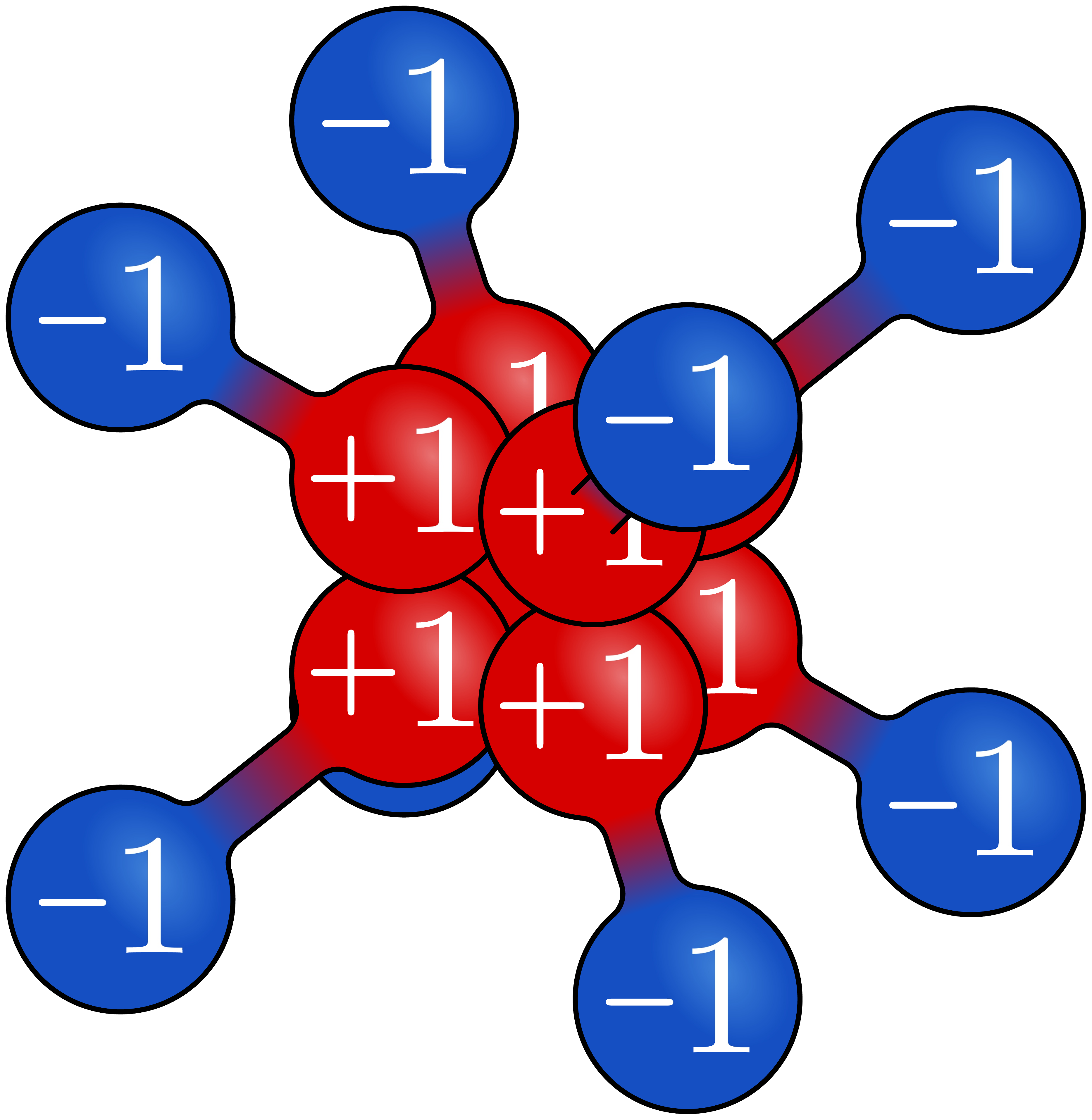}
    \caption{Structure of the three-dimensional gate whose characteristic's zeros ($\chi(\vec{k})=0$) are shown in Fig.~\ref{fig:ffig4}b. It is constructed from two Eq.~\eqref{eq:Hex} gates superimposed on the diagonal planes.}
    \label{fig:fig3D}
\end{figure}

Finally, we turn to a 3D model, which is a natural generalization of Eq.~\eqref{eq:Hex}. The corresponding gate acts on a $4\times4\times4$ cube, moving charges between the inner 8 sites and the outer corners; this is illustrated in Fig.~\ref{fig:fig3D}. The resulting characteristic equation is given by
\begin{align}
   \nonumber& \chi(\Vec{k})=16\cos\klam[\Big]{\frac{k_x}{2}}\cos\klam[\Big]{\frac{k_y}{2}}\cos\klam[\Big]{\frac{k_z}{2}} \times \nonumber \\& \times \big[4\cos(k_x)\cos(k_y)\cos(k_z) - \nonumber \\ \nonumber &-2\bigl(\cos(k_x)\cos(k_y)+\cos(k_y)\cos(k_z)+\cos(k_x)\cos(k_z)\big)\\&+\cos(k_x)+\cos(k_y)+\cos(k_z)-1\big]\nexp{\x\frac{k_x+k_y+k_z}{2}}.
\end{align}
The zeros of this function were sketched in Fig.~\ref{fig:ffig4}b of the paper. 

\section{Decay of Correlations in 2D Models}

Here we discuss how the presence of conserved momentum modes in 2D affects the decay of the spin-spin autocorrelation. As we argued in the main text, the long time behavior is dominated by the density of states at low frequencies, via Eq.~\eqref{eq:C_from_DOS}, which can be determined from the dispersion relation $\omega(\vec{k}) \sim |\chi(\vec{k})|^2$. We first discuss the details of this calculation for the model~\eqref{eq:Hex} and then provide some general considerations applicable to arbitrary dispersion relations. 

\subsection{Correlations in Model~$\boldsymbol{\eqref{eq:Hex}}$}

The dispersion relation is
\begin{align}\label{eq:disprel_2D}
    \omega(k_x,k_y) \sim |\chi(k_x,k_y)|^2 = \cos^2(k_x/2) \cos^2(k_y/2) \times \nonumber \\
    \times \left[\cos(k_x) + \cos(k_y) - 2\cos(k_x)\cos(k_y) \right]^2.
\end{align}
As already noted, it vanishes at the origin as well as along the lines shown in Fig.~\ref{fig:ffig2}b. To evaluate the DOS we calculate $N(\omega)$, the number of states in the frequency range $[0,\omega]$ (i.\,e., $N(\omega)$ is the area of the region in the BZ delineated by the condition $\omega(k_x,k_y) \leq \omega$) and then take a derivative, $\rho(\omega) = \frac{\mathrm{d}N(\omega)}{\mathrm{d}\omega}$. We split up $N(\omega)$ into a sum of contributions from different regions of the BZ which we evaluate independently. 

\paragraph{1) $\vec{k} \approx \vec{0}$}--- Near the origin, the dispersion is dictated by the multipole symmetries alone; it has an expansion $\omega(\vec{k}) \approx k^4 + \mathcal{O}(k^5)$ where $k \equiv \lVert\vec{k}\rVert$. This is spherically symmetric and has the sub-diffusive scaling that one expects based on the fact that the model conserves dipole moments~\footnote{Note that while the model conserves certain components of the quadrupole moment, it does not conserve all; if it did, it would have an even lower power $\omega \sim k^6$.}. This gives $N(\omega) \sim (\omega^{1/4})^{2} \sim \omega^{1/2}$.% and a DOS $\rho(\omega) \sim w^{-1/2}$. 

\paragraph{2) $k_x,k_y \approx \pi$}--- At the corner of the BZ, the two lines of conserved momenta $k_x=\pi$ and $k_y=\pi$ meet. Expanding around this point, we find $\omega \approx \delta k_x^2 \delta k_y^2$, where $k_i = \pi - \delta k_i$. This situation is similar to the case of U$(1)$ subsystem symmetries that was considered in Ref. \cite{Iaconis_2019}, which has the same dispersion near the origin. This gives rise to a logarithmic correction to the DOS: $N(\omega) \sim \omega^{1/2} \log(\omega)$ (we will re-derive the logarithmic correction below).

\paragraph{3) $k_x \approx \pi$, $k_y \approx k^*$}--- There is another similar point where the $k_x=\pi$ line of conserved momenta crosses the loop of conserved modes; this happens when $k_y = k^*$ where $\cos{k^*} = 1/3$. In fact, there are four such points in total, $(k_x,k_y) = (\pi,k^*), (\pi,-k^*),(k^*,\pi),(-k^*,\pi)$, each with identical contributions to the DOS. Expanding around one of these points, we again find $\omega(\vec{k}+\delta\vec{k}) \approx \delta k_x^2 \delta k_y^2$, so these also lead to the same contribution, $N(\omega) \sim \omega^{1/2} \log(\omega)$.

\paragraph{4) $k_x \approx \pi$, $k_y \not\approx k^*, \pi$}--- Along the line $k_x = \pi$, but away from the aforementioned crossing points, we have $\omega \sim \delta k_x^2$ (with a prefactor that depends on $k_y$ but is finite everywhere in this regime). $N(\omega)$ is therefore approximately the area of a rectangle with one side of length $\mathcal{O}(\omega^{1/2})$ and the other side of length $\mathcal{O}(1)$, which gives $N(\omega) \sim \omega^{1/2}$. Same for the line $k_y \approx \pi$. 

\paragraph{5) The loop}--- Intuitively, the contribution of the loop is similar to the previous case: $N(\omega)$ counts the area of a `fattened' loop, extended to a size $\sim \omega^{1/2}$ in the direction transverse to it. and one expects a contribution of size $N(\omega) \sim \omega^{1/2}$. A more detailed calculation confirms this expectation.

\paragraph{Putting it together}--- As we saw, most of contributions we considered scale to leading order st small $\omega$ as $N(\omega) \sim \omega^{1/2}$. The exception are the five points where different lines of conserved momenta cross, at which $N(\omega)$ has an additional logarithmic enhancement. These latter contributions dominate at the smallest frequencies, leading to $N(\omega) \sim \omega^{1/2}\log(\omega) + \mathcal{O}(\omega^{1/2})$. Taking a derivative, we get $\rho(\omega) \sim \omega^{-1/2}\log(\omega) + \mathcal{O}(\omega^{-1/2})$  and plugging this into Eq.~\eqref{eq:C_from_DOS} gives an autocorrelation $C(\vec{0},t) \sim t^{-1/2}\log(t) + \mathcal{O}(t^{-1/2})$.

\subsection{General Considerations}

As we saw in Fig.~\ref{fig:ffig4}a, and in the previous section, there are many other examples of conserved momentum modes arranged along various shapes in the BZ. While the details of these shapes should show up in the spatial structure of $C(\vec{r},t)$, the $\vec{r}=0$ autocorrelation is dictated by a few relevant features that enter into the calculation of $\rho(\omega)$. Similarly to the example above, we will split up the calculation of $\rho(\omega)$ into two contributions: continuous lines and singular points (such as crossing points between two lines).

Along the lines, $N(\omega)$ is the area of a strip given by broadening the line to include points with $\omega(\vec{k}) \leq \omega$; The width of this strip depends on the expansion of $\omega$ along the transverse direction. Taylor expanding around a point $\vec{k}$ on the contour, we can write the leading order term as $\chi(\vec{k}+\delta\vec{k}) \approx (a\mathop{}\!\delta k_x + b\mathop{}\!\delta k_y)^m$, where $a,b \in \mathbb{R}$ generically depend on $\vec{k}$. We call $m$ the {\it multiplicity} of the line: the simplest possibility, realized in the model~\eqref{eq:Hex}, is $m=1$ which gives $N(\omega)\sim \omega^{1/2}$ and thus $C(\vec{0},t)\sim t^{-1/2}$. However, higher multiplicities are possible, as in the model shown on the left of Table~\ref{tab:Table1}, where near the $k_x=0$ axis, we have $\omega(\vec{k}) \sim k_y^4$ (and vice versa). In general, we then get $N(\omega) \sim \omega^{1/2m}$ which leads to a contribution $C(\vec{0},t) \sim t^{-1/2m}$ in the autocorrelation. 

An enhanced contribution to the DOS can arise from {\it singular points}, where the leading order Taylor expansion of $\chi(\vec{k})$ does not have the form $(a\mathop{}\!\delta k_x + b\mathop{}\!\delta k_y)^m$. This can occur for various reasons: at isolated points, at a meeting point of multiple branches of the curve, or if the form of the expansion changes at a point along the curve. In our list of singular points, we also include points where two or more lines touch: while in this case we do have an expansion of the above form, with $m>1$, the value of $m$ changes discontinuously as we move away from the touching point. 

Without loss of generality, we can write the leading term in the Taylor expansion of the characteristic function around a point as~\cite{fulton1989algebraic}
\begin{align}\label{eq:SingPoints}
    \chi(\vec{k}+\delta\vec{k}) \approx \prod_i (a_i\mathop{}\!\delta k_x + b_i\mathop{}\!\delta k_y)^{m_i},
\end{align}
where the coefficients $a_i, b_i$ might be complex in general. When $a_i/b_i$ is real, we can picture the corresponding term as the tangent of the curve at $\vec{k}$, each appearing with some multiplicity $m_i$. $m_i > 1$ might occur because the line in question itself has a non-trivial multiplicity, or because two different lines share the same tangent, i.e. when they have a touching point at $\vec{k}$. Terms where $a_i/b_i$ is complex are singularities that do not arise from the meeting of contour lines; an extreme example is an isolated singular point (such as $\vec{k}=\vec{0}$ for Eq.~\eqref{eq:2dchi}) where all tangent lines are complex. For another example of complex roots, consider $\vec{k}=\vec{0}$ in the left panel of Table~\ref{tab:Table2}.

\begin{figure}
  		\centering
  		\includegraphics[width=0.85\linewidth]{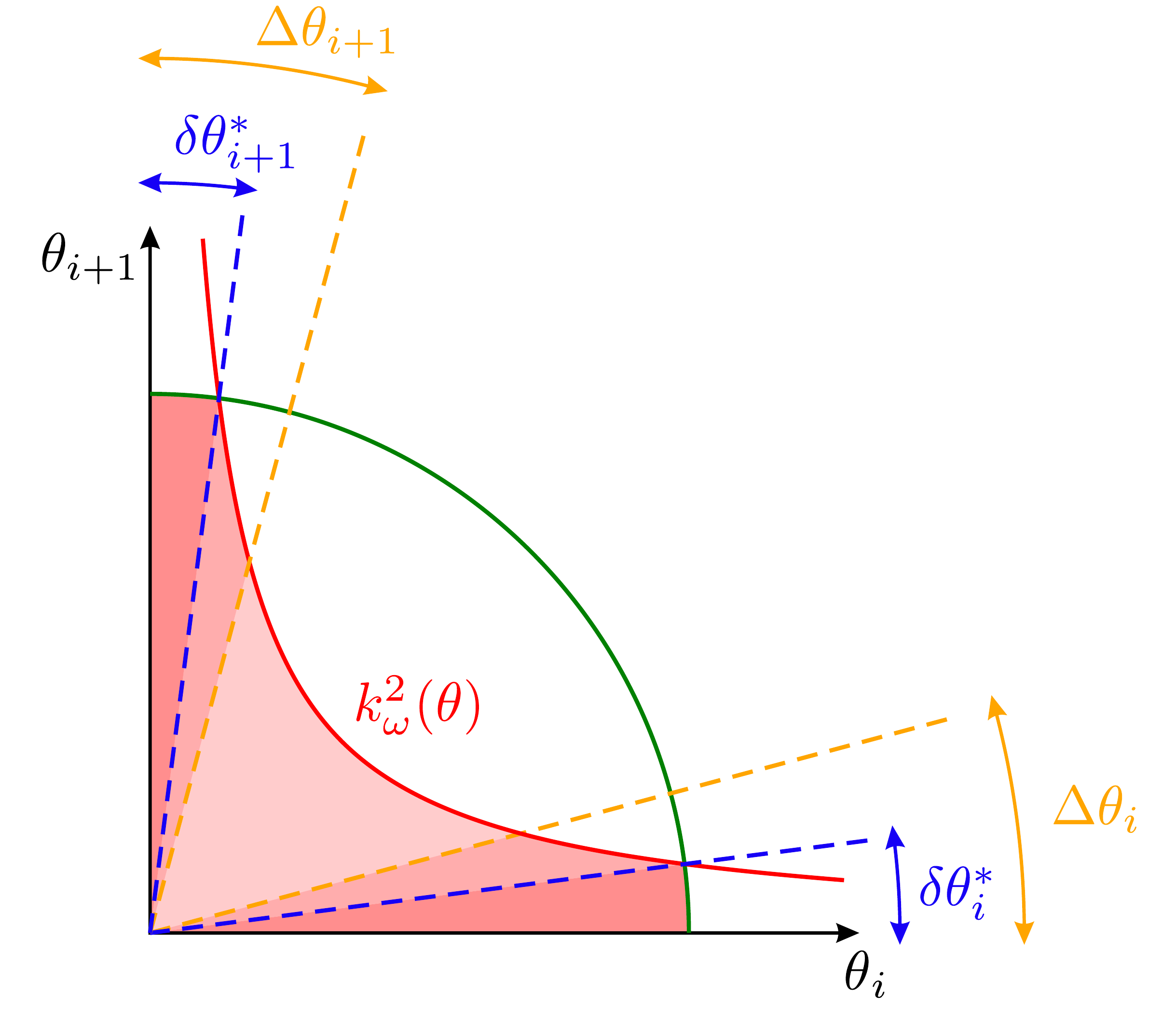}
  		
  		\caption{Evaluating the integral over angles in Eq.~\eqref{eq:theta_int}. We evaluate a contribution from the region between two subsequent zeros of $f(\theta)$, $\theta_{i}$ and $\theta_{i+1}$ which we take to be the $x$ and $y$ axes for simplicity. We are interested in the area bounded by the curve $k^2_\omega(\theta)$ (red line) which diverges at $\theta_{i,i+1}$; to regularize we restrict to within a circle of radius $\Delta k$ (green line). This splits the area into three parts, shown in different shades of red.}
  		
  		\label{fig:theta_int}
  	\end{figure}

Using polar coordinates, $(\delta k_x,\delta k_y) \equiv (k \cos{\theta}, k \sin{\theta})$ we can rewrite Eq.~\eqref{eq:SingPoints} as $\chi(\vec{k} + \delta \vec{k}) \approx k^{m} f(\theta)$, where $m \equiv \sum_i m_i$ is the total multiplicity of the singular point. Consider now the set of points defined by the condition $\omega(k,\theta) \sim |\chi(k,\theta)|^2 = k^{2m} |f(\theta)|^2 = \omega$; this is equivalent to $k = k_\omega(\theta) = \omega^{1/2m} |f(\theta)|^{-1/m}$. The state counting then becomes
\begin{align} \label{eq:theta_int}
    N(\omega) &\sim \int_{\omega(\vec{k}) \leq \omega}\mathrm{d}^2k = \nonumber \\
    &= \int_{0}^{2\pi} \mathrm{d}\theta \int_{0}^{k_\omega(\theta)}\mathrm{d}k\!\mathop{} k = \omega^{1/m} \int_{0}^{2\pi} \mathrm{d}\theta\!\mathop{} |f(\theta)|^{-2/m}.
\end{align}
The integrand in Eq.~\eqref{eq:theta_int} diverges along the tangent lines, $\theta = \theta_i$, defined by $a_i \cos{\theta_i} + b_i \sin{\theta_i} = 0$. We can split the integral into regions close to these angles and regions away from them, as illustrated in Fig.~\ref{fig:theta_int}; for the latter, the integrand is finite and we are simply left with a contribution that scales as $N(\omega) \sim \omega^{1/m}$.

The contribution from regions close to $\theta \approx \theta_i$ depends on how $f(\theta)$ vanishes at this point, which is set by the multiplicity $m_i$: $f(\theta_i + \delta\theta) \sim \delta\theta^{m_i} + \mathcal{O}(\delta\theta^{m_i+1})$. Close to the tangent line, we thus have $\int_{\theta_i}^{\theta_i+\Delta\theta_i}\mathrm{d}\theta\!\mathop{} (\theta-\theta_i)^{-2m_i/m}$ (here $\Delta\theta_i$ is some small angle beyond which the leading order Taylor expansion is no longer valid) %; e.g. if $f(\theta_i + \delta\theta) = \alpha \mathop{}\!\delta\theta^{m_i} + \beta \mathop{}\!\delta\theta^{m_i+1} + \ldots$ then we can take $\Delta\theta_i \approx \alpha/\beta$). 
If $2 m_i < m$ then the integral converges and we are left with $N(\omega) \sim \omega^{1/m}$. If $2 m_i \geq m$ it diverges at $\omega\to 0$ and we need to regularize it.

To regularize the integral, note that our initial Taylor expansion of $\chi(\vec{k})$ is only valid within some neighborhood of the singular point we consider. In particular, we should restrict ourselves to a disc of some small radius, $k \leq \Delta k$, around this point and count low-frequency states only within this circle. This means that the part of the integral~\eqref{eq:theta_int} near $\theta_i$ should be further decomposed into two parts: those $\theta$ for which $k_\omega(\theta) \leq \Delta k$ and those where $k_\omega(\theta) > \Delta k$; this is again illustrated in Fig.~\ref{fig:theta_int}. The point separating the two cases is defined by $\Delta k^{2m} |f(\theta_i+ \delta\theta_i^*)|^2 \sim \Delta k^{2m} \left(\delta\theta_i^*\right)^{2m_i} = \omega$ which implies $\delta\theta_i^* \sim \omega^{1/2m_i}/\Delta k^{m/m_i}$. 
We thus replace the contribution from $\theta\approx\theta_i$ with a sum of two terms: \begin{align}
   \int_{\theta_i}^{\theta_i+\Delta\theta_i} k_\omega^2(\theta) \text{d}\theta \to \Delta k^2 \int_{\theta_i}^{\theta_i+\delta\theta_i^*} \text{d}\theta + \int_{\theta_i+\delta\theta_i^*}^{\theta_i+\Delta\theta_i} k_\omega^2(\theta) \text{d}\theta.
\end{align}
The first is simply the area of the circular segment between $\theta_i$ and $\theta_i + \delta\theta_i^*$, which gives $N(\omega) \sim \Delta k^2 \delta\theta_i^* \sim \omega^{1/2m_i}$. The second behaves differently depending on whether $2m_i=m$ or $2m_i > m$:
\begin{align*}
    %\int_{\delta\theta_i^*}^{\Delta\theta} k_\omega^2(\theta) \sim
    \omega^{\frac{1}{m}} \int_{\delta\theta_i^*}^{\Delta\theta} \frac{\text{d}\theta}{\theta^{2m_i/m}} \sim \begin{cases} \omega^{\frac{1}{m}} \log(\omega) \ \quad &2m_i=m \\ \omega^{\frac{1}{2m_i}} + \mathcal{O}(\omega^{\frac{1}{m}}) \quad &2m_i>m \end{cases}.
\end{align*}

To summarize we find three distinct cases:
\begin{itemize}
    \item if $2m_i<m$ then $N(\omega) \sim \omega^{1/m}$,
    \item if $2m_i=m$ then $N(\omega) \sim \omega^{1/m}\log(\omega)$,
    \item if $2m_i>m$ then $N(\omega) \sim \omega^{1/2m_i}$,
\end{itemize}
to leading order. Remembering that $m=\sum_i m_i$, the condition $2m_i > m$ is equivalent to $m_i > \sum_{j\neq i} m_j$, such that one tangent line dominates over all the others.

The first of these possibilities is realized at the origin in Fig.~\ref{fig:ffig4}a and in the right panel of Tables~\ref{tab:Table1} and~\ref{tab:Table2}. In these cases $m$ is simply the number of lines that cross; for example, the dispersion of Fig.~\ref{fig:ffig4}a leads to $C(\vec{0},t) \sim t^{-1/4}$. 
The second case was realized at the crossing points in the model~\eqref{eq:Hex}, with $m_1=m_2=1$. This is also the case near the origin in the left figure of Table~\ref{tab:Table1} where $m_1=m_2=2$. 
In some sense, the most interesting is the last possibility. In this case, the DOS has an entirely different power law than what the naive dimension counting $\omega \sim k^{2m}$ would suggest; consequently correlations have a slower decay $C(\vec{0},t) \sim t^{-1/2m_i}$ A simple situation where this occurs is a point where two lines have a touching point. 

\section{Details on the implementation of cellular automaton dynamics}

As discussed in the main text, at each application of a local gate we choose randomly (with probabilities $1/3$ each) between three possibilities: applying $G_{\vec{x}}$, applying its inverse, or doing nothing. 
In the first two cases, the update is applied only if it does not lead to a violation of the local constraint $|s_{\vec{r}}| \leq S$ on any site, otherwise we leave the configuration unchanged.
     %
%We employ a stochastic (blocked) cellular automata circuit built from local gates $G_x$ of range $(2\ell+1)$, respecting the set of modulated symmetries we discussed in the main text.
%
%As explained there, these local gates $G_x$ are specified by a set of integers $\{n_i\}$, such that when applying $G_x$, the spins are updated as $s_{x+i}\to s_{x+i}\pm n_i$ with $i\in \{-\ell,\dots,\ell\}$. 
%
These updates are randomly applied among those configurations for which $|s_{x+i}\pm n_i| \leq S$, such that the corresponding transition rates between two different local configurations $\vec{s},\vec{s}^{\prime}$ are symmetric, i.e., $\gamma_{\vec{s}\to \vec{s}^{\prime}}=\gamma_{\vec{s}^{\prime}\to \vec{s}}$.
This ensures that detailed balance is satisfied with respect to the ``infinite temperature'' (uniformly random) ensemble, which is therefore a stationary state of this stochastic process.
Note that this implementation differs from that used in previous works~\cite{Morningstar_2020,Feldmeier_2020,Iaconis_2019,Iaconis_2021}, where all local updates consistent with symmetry requirements were allowed.
%
%Let us consider three local spin configurations $\vec{s}_1,\vec{s}_2$ and $\vec{s}_3$ on the same sites $\{x-\ell,\dots,x+\ell\}$ where a gate $G_x$ acts upon, and sharing the same value of the conserved quantities.
%
%Our implementation only allows (symmetric) transitions given by $s_{x+i}\to s_{x+i}\pm n_i$, and leaves the configuration invariant whenever the condition  $|s_{x+i}\pm n_i| \leq S$ is not fulfilled.
%
%Instead, previous works~\cite{Morningstar_2020,Iaconis_2021} allowed any possible transition among these three configurations with equal probability.
%
%For example, transitions among any set of configurations with the same charge and dipole moment.
%
Thus, certain direct transitions in the latter implementation require multiple updates in the former.
%

%At each time step we randomly pick an integer $m\in \{-\ell,\dots,\ell\}$ and apply all gates on sites $x = m \text{ (mod } 2 \ell+ 1)$, such that gates lie adjacent to each other, but without overlapping. 
At each time step, we randomly pick a non-overlapping complete covering of the (1D or 2D) lattice by the gates $G_{\vec{x}}$.
For a model in 1D with gates acting on $2\ell$ sites, we pick randomly an integer $m\in \{,\dots,\ell-1\}$ and apply all gates on sites $x = m \text{ (mod } \ell)$. 
Similarly, in 2D with gates of size $\ell\times\ell$ we pick two integers $m_x,m_y \in \{0,\ldots,\ell-1\}$ and shift the gates accordingly. 
%
%An evolution till time $t$ then consists on the application of $t$ consecutive layers.
%
Moreover, for periodic boundary conditions we choose system sizes that are multiple of $\ell$. 

\section{Construction of longer-range gates}
  	
  	In the main text, we introduced the notion of modulated symmetries considering size-$3$ gates; their quantum versions are Hamiltonians with 3-site interactions. 
  	However, one does not need to restrict the analysis to this particular size and in fact, one can easily construct longer-range terms sharing the same sets of symmetries.
  	Such construction follows the ideas of Ref.~\onlinecite{Feldmeier_2020}, where they were used to obtain models with $m$th-moment conservation, from those which only conserve the $(m-1)$th moment.
  	
  	Consider the family of size-$3$ gates, $G^{(q,p)}$ in Eq.~\eqref{eq:H1d}, determined by the strings of numbers  $\{n_i\}=(\mp q,\pm p,\mp q)$ acting on sites $(x,x+1,x+2)$ with associated characteristic equation $q - p r + qr^2=0$.
  Let us start constructing range-$4$ gates from range-$3$ ones and address the general case afterwards.
  Adding up strings corresponding to two overlapping range-$3$ gates we find
 \begin{align}
 \begin{array}{r}
     (-q,+p,-q,0)\\ 
     +(0,-q,+p,-q) \\[0.5ex] \hline
     (-q,p-q,p-q,-q)
\end{array},
 \end{align}
 such that the resulting gate $G_x=\{n_i\}=(-q,p-q,p-q,-q)$ acts on sites $(x,x+1,x+2,x+3)$.
 The associated recurrence relation takes the form
  \begin{equation} \label{eq:eqreclong}
     -q\alpha_j +(p-q)\alpha_{j+1} +(p-q)\alpha_{j+2} -q\alpha_{j+3}=0.
 \end{equation}
 with characteristic equation 
 \begin{equation} \label{eq:characlong}
     -q +(p-q)r +(p-q)r^2 -qr^3=0.
 \end{equation}

 Generically, such a combination results in a characteristic equation which contains additional roots to those we started from. 
 Indeed, due to the linearity of Eq.~\eqref{eq:eqreclong}, these include the roots of $-q +pr-qr^2$, i.e., independent solutions corresponding to each of the original gates, as can be clearly seen by factorizing Eq.~\eqref{eq:eqreclong} in the form
\begin{equation}
    (-q+pr-qr^2)(r+1)=0.
\end{equation}

In this case, we find the additional solution $r=-1$ (i.\,e., staggered modulation). However, a cellular automaton (or quantum Hamiltonian) evolution including range-$3$ but also range-$4$ gates, will only host those symmetries corresponding to the common set of solutions, i.\,e. those for which $-q +pr-qr^2=0$. 

In general, if two gates $\{n_{\vec{r}^{\prime}}'\}, \{n_{\vec{r}^{\prime \prime}}''\}$ 
%with recurrence relations $\{\alpha^{\prime}_{\vec{r}^{\prime}+\vec{j}}\}$, $\{\alpha^{\prime \prime}_{\vec{r}^{\prime \prime}+\vec{j}}\}$,
have a common symmetry $\mathcal{Q}$,  their element-wise combination, $\{n_{\vec{r}}\} \equiv \{n'_{\vec{r}} \pm n''_{\vec{r}}\}$, also features the same symmetry.
%h
Indeed, the resulting recurrence equation $\sum_{\vec{r}}(n'_{\vec{r}}\pm n''_{\vec{r}})\alpha_{\vec{r}}=0$ is satisfied if both
$\sum_{\vec{r}} n'_{\vec{r}} \alpha_{\vec{r}}$ and $\sum_{\vec{r}} n''_{\vec{r}} \alpha_{\vec{r}}$
%$\{\alpha^{\prime}_{\vec{r}^{\prime}+j}\}$, $\{\alpha^{\prime \prime}_{\vec{r}^{\prime \prime}+j}\}$
vanish.

Nevertheless, additional solutions might arise from a cancellation between the two terms, as it was the case of our previous example.

This construction allows to numerically address the universal properties of systems with a set of conserved quantities, avoiding the possibility of strong Hilbert space fragmentation appearing for small spin representations $S$.

\begin{figure}
    \centering
    \includegraphics[width=\linewidth]{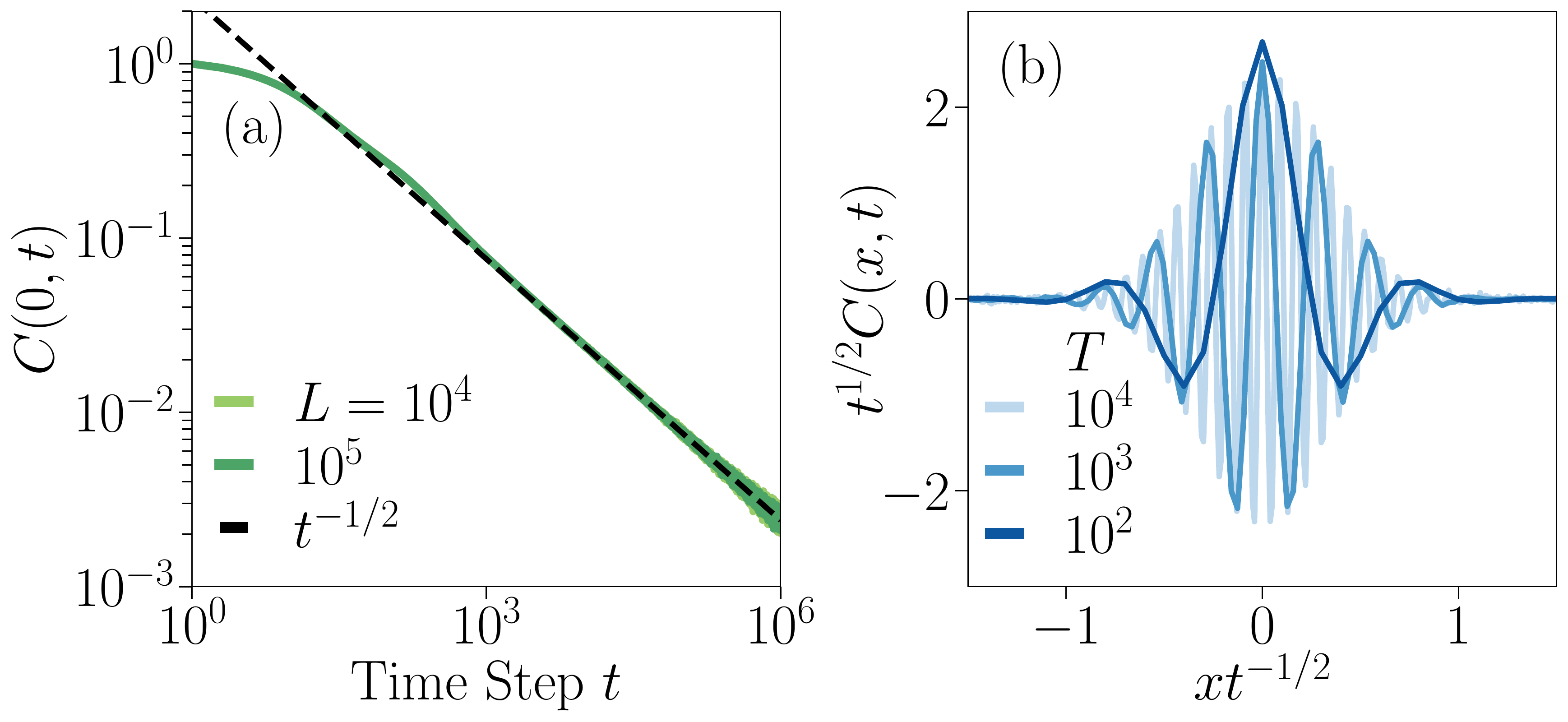}
    \caption{Evolution combining size-$3$ and size-$4$ gates with $(p,q)=(3,2)$. }
    \label{fig:fig6}
\end{figure}

Fig.~\ref{fig:fig6} shows numerical data for a stochastic evolution combining layers of range-$3$ gates (as in Eq.~\eqref{eq:H1d}) and with layers of range-$4$ ones, and $(p,q)=(3,2)$. The result agrees with data shown in Fig.~\ref{fig:ffig1}.

\section{Mazur Bound for exponential localized symmetries in 1D}

  Similar to Refs.~\onlinecite{Rakovszky_2020,Moudgalya_2021}, one can use the Mazur bound~\cite{MAZUR1969533} to lower bound the long-time average of the charge autocorrelation at the boundaries in the case with exponentially localized symmetries. In the following, we consider the left boundary correlator $ \avg{s_{\ell}(t)s_{\ell}(0)}$, although the exact same result applies to the right one. Given a set of conserved quantities $\{\mathcal{Q}_{\alpha}\}$, the Mazur bound states that
  \begin{align} \nonumber
  C_{\ell}&\equiv \lim_{T\to \infty}\frac{1}{T}\int_0^T \mathrm{d}t\!\mathop{}\avg{s_{\ell}(t)s_{\ell}(0)} \\
  &\geq \sum_{\alpha,\beta} \avg{s_{\ell},Q_{\alpha}}(K^{-1})_{\alpha,\beta}\avg{Q_{\beta},s_{\ell}}\equiv M_{\ell},
  \end{align}
  where $\avg{\cdot,\cdot}$ denotes the Frobenius inner product of two observables, and $K$ is the Gram-Schmidt matrix with elements $K_{\alpha,\beta}=\avg{\mathcal{Q}_\alpha, \mathcal{Q}_\beta}$. This formula simplifies whenever $\{\mathcal{Q}_{\alpha}\}$ form an orthogonal set with respect to $\langle \cdot,\cdot \rangle$ such that $K_{\alpha,\beta}\propto \delta_{\alpha,\beta}$. In our case there exists (at least) two such conserved quantities corresponding to the exponentially localized U$(1)$'s whose support is localized at the two boundaries of the system. Let us denote by $\mathcal{Q}_{\ell}$ ($\mathcal{Q}_r$) the charge localized at the left (right) boundary of the system with boundary conditions $\alpha^{\ell}_0=1, \alpha^{\ell}_{L-1}=0$ $(\alpha^r_0=0, \alpha^r_{L-1}=1)$. Their exact form is given by
  \begin{align} \nonumber
      &\alpha^{k=\ell,r}_x = \frac{1}{r^{L-1}-r^{-(L-1)}}\left[(\alpha^k_{L-1}-\alpha^k_0 r^{-(L-1)})r^x \right.\\
     \label{eq:alpha_x} & \left.+ (\alpha^k_0 r^{L-1} - \alpha^k_{L-1})r^{-x}) \right],
  \end{align}
  with $r>1$ the largest root of the associated characteristic equation.
  Although we can get exact formulas for a finite system of size $L$, we would rather focus on the scaling in the limit $L\to \infty$, where Eq.~\eqref{eq:alpha_x} takes the asymptotic form 
   \begin{align} \nonumber
  	\alpha^{\ell}_x \approx r^{-x},\,\textrm{ and}\,\,\,\,\,\, \alpha^{r}_x \approx r^{x-(L-1)}.
  \end{align}

  Thus, one finds $\avg{\mathcal{Q}_{\ell},\mathcal{Q}_r}\approx \frac{L}{r^{L}}\to 0$, i.\,e., $\mathcal{Q}_{\ell},\mathcal{Q}_r$ become orthogonal in the thermodynamic limit, simplifying the expression for the bound.
  Together with $\avg{\mathcal{Q}_{\ell},\mathcal{Q}_{\ell}},\avg{\mathcal{Q}_r,\mathcal{Q}_r}\approx \avg{(s_x)^2}\frac{r^2}{r^2-1}$, Mazur bound becomes
  \begin{align} \nonumber
      &M_{\ell} \approx \frac{\avg{s_{\ell},\mathcal{Q}_{\ell}}^2}{\avg{\mathcal{Q}_{\ell},\mathcal{Q}_{\ell}}} +  \frac{\avg{s_{\ell},\mathcal{Q}_r}^2}{\avg{\mathcal{Q}_r,\mathcal{Q}_r}} \\
      &=\avg{(s_x)^2}\frac{r^2 -1}{{r^2}}  \left[ (\alpha_0^{\ell})^2+  (\alpha_0^r)^2 \right]= \avg{(s_x)^2}\frac{r^2 -1}{r^2},
  \end{align}
  with $\avg{(s_x)^2}=S(S+1)/3$ the infinite temperature expectation value of $(s_x)^2$. This implies that the presence of exponentially localized symmetries, leads to infinitely long-lived correlations at the boundaries of a 1D system. For the case studied in the main text $r=(3+\sqrt{5})/2$. This value corresponds to the black dashed line in the inset of Fig.~\ref{fig:ffig1}b. 
  
  	\section{Counting the number of independent modulated symmetries}
  	
  We would like to count the number of independent modulated symmetries corresponding to independent solutions of the two-dimensional recurrence equation \eqref{eq:rec2D2_main}.  
  To do so, let us consider a finite system of linear size $L$ and open boundary conditions (OBC), such that all solutions of the recurrence relation correspond to exact symmetries.
  Imposing periodic boundary conditions (PBC), leads to the additional constraints $\alpha_{i+L,j}=\alpha_{i,j+L}=\alpha_{i,j}$ for all $i,j$, which then depends on the system size. 

  Let us first consider a particular example, and only later extend the resulting counting to the general case. Consider the recurrence relation
\begin{equation} \label{eq:rec2D_Htilt}
 4\alpha_{i,j} - \alpha_{i-1,j}-\alpha_{i+1,j}-\alpha_{i,j-1} -\alpha_{i,j+1} = 0,
\end{equation}
which is a two-dimensional second-order linear equation in both $x$ and $y$ directions.
This e.g., is the associated equation to the set of $3\times 3$ local gates
\begin{align}
    \nonumber G&= \{n_{0,0},n_{-1,0},n_{1,0},n_{0,1},n_{0,-1}\} \\
    &= \{4,-1,-1,-1,-1\},
\end{align}
which corresponds to our 2D example of a system with (only) exponential localized symmetries.

Obtaining an analytical exact solution for a 2D recurrence relation, requires either to obtain the corresponding generating function~\cite{10.2307/2984940}, or rewriting the system as a Sylvester equation~\cite{10.2307/3215145}. In both cases, and even with the full solution at hand, it is still rather involving to extract information from it. 
Alternatively, one can recursively solve $\alpha_{i,j}=f(\alpha_{m,n})$ with $m\neq i,n\neq j$, after fixing a minimal set of initial (or boundary~\cite{10.2307/3215145}) conditions.
%
%In 1D, one can equivalently solve the associated characteristic equation, or in general, write such equation as a system of linear equations and find the complete solution~\cite{10.2307/3215145}. 
%%
%One possible way of writing such a recursion corresponds to
%\begin{equation} \label{eq:rec2D_1}
% 4\alpha_{i,j} = \alpha_{i-1,j}+\alpha_{i+1,j}+\alpha_{i,j-1} +\alpha_{i,j+1},
%\end{equation}
%which requires to fix the value along the first two rows, i.\,e., $\alpha_{i,0}$ and $\alpha_{i,1}$. 
%
%Analogously, along direction $x$, the state on a given site $(i,j)$ depends on the two adjacent columns $i-1,i+1$ and thus we will fix the initial values along the first and last columns $\alpha_{0,j}$ and $\alpha_{L,j}$.  
%
%This means that we would need to fix $D\equiv 4L -4$ initial (boundary) values. 
%
For example, Eq.~\eqref{eq:rec2D_Htilt} can be expressed as
\begin{equation} \label{eq:rec2D2}
	\alpha_{i+1,j}=4\alpha_{i,j} - \alpha_{i-1,j}-\alpha_{i,j+1}-\alpha_{i,j-1},
\end{equation}
such that it is sufficient to fixed the values along the first two left columns $\alpha_{0,j},\alpha_{1,j}$, and a single row at the bottom of the lattice $\alpha_{i,0}$, to obtain the value of any $\alpha_{i,j}$ corresponding to an exact conservation law.
This implies that one needs to fix $D=3L-2$ initial conditions, corresponding to the entries of $\alpha_{0,j},\alpha_{1,j}$ and $\alpha_{i,0}$.
In general, given a $n$th-order two dimensional recurrence relation, one needs to fix $D=\mathcal{O}(L)$ values.

However, not every choice of initial conditions corresponds to a linearly independent conserved quantity.
Let us collect $\alpha_{0,j},\alpha_{1,j}$ and $\alpha_{i,0}$ into a vector $\vec{v}_0$ with $D$ entries.
Then, any choice of initial conditions can be written as a linear combinations of elements of the canonical basis $e_i$ with zeros in every entry except the $i$th one.
This was also the case in 1D: There we encountered a second-order recurrence relation which requires fixing $(\alpha_0,\alpha_1)$ (or equivalently $(\alpha_0,\alpha_{L-1})$).
Thus, the set of initial conditions is a two-dimensional vector space with canonical basis vectors: $\vec{e}_1\equiv (1,0)$ and $\vec{e}_2\equiv(0,1)$. 

In 2D, this implies that there exist a subextensive number of linearly independent conserved quantities $D=\mathcal{O}(L)$.
Indeed, this scaling is consistent with the fact that factorizable solutions of the form $\alpha_{i,j}=(r_x)^i (r_y)^j$, lead to one-dimensional manifolds parametrized by $r_x, r_y$.
An example of this was shown in Fig.~\ref{fig:ffig2}b and Fig.~\ref{fig:ffig4}, but it also holds for the exponential localized solutions of Eq.~\eqref{eq:rec2D_Htilt}.

We emphasize that for the previous computation any  such choice corresponds to an exact symmetry.
Nevertheless, in the case of PBC, neither exponential symmetries nor most modes are exactly realized; moreover, we also note that if $\vec{k}$ is a conserved mode, so is $-\vec{k}$.

\section{Symmetries in real space}

In the main text, we found that the spatial staggered subsystem symmetries $\mathcal{S}_x,\mathcal{S}_y$ of the 2D model \eqref{eq:Hex}, correspond to the inverse Fourier transform of solutions lying along the $k_x,k_y=\pi$ lines in momentum space
\begin{align} 
    &\mathcal{S}_{y_0} \equiv\sum_{x}(-1)^xs_{x,y_0} =\frac{1}{L}\sum_{k_y} \mathrm{e}^{-\mathrm{i}y_0k_y}{\mathscr{s}}_{\pi,k_y},\\
    &\mathcal{S}_{x_0}\equiv \sum_{y}(-1)^ys_{x_0,y} =\frac{1}{L}\sum_{k_x} \mathrm{e}^{-\mathrm{i}x_0k_x}{\mathscr{s}}_{k_x,\pi}.
\end{align}

These can be written in the following form
\begin{align}
    &\mathcal{S}_{y_0}=\frac{1}{L^2}\sum_{x,y}\alpha(\vec{r};y_0)s_{x,y},
\end{align}
(analogous for $\mathcal{S}_{x_0}$) with the modulation given by the inverse Fourier transform of the function $f(\vec{k})=\mathrm{e}^{-\mathrm{i}k_yy_0}$
\begin{align}
    \alpha(\vec{r})=\sum_{\mathclap{\vec{k} \in \{(k_x=\pi,k_y)\}}}\mathrm{e}^{\mathrm{i}\vec{k}\cdot \vec{r}}f(\vec{k}),
\end{align}
taking values along the lines $k_x=\pi$ ($k_y=\pi$). 
Without loss of generality, and evaluating at $y_0=0$, this implies that subsystem symmetries appear as the ``equally weighted'' inverse Fourier transform along the one-dimensional manifold of solutions of $\chi(\vec{k})=0$ corresponding to $k_x=\pi$.
This result is quite natural: Along these straight lines, one finds the most localized object in real space (i.e., a Kronecker delta) via the 1D inverse Fourier transform of the constant function.

A natural question is whether one could similarly construct a subsystem symmetry from the inverse Fourier transform of conserved modes along some closed loop in the BZ, such as the ones we encountered in this paper. Here, we will argue that this is not the case: the quantities that can be constructed in this case decay asymptotically as $\|\vec{r}\|^{-1/2}$ at large distance along almost all directions in space, implying that they are spread out around the entire system. 

To build some intuition, consider a simple example where the conserved momenta are along a circle in momentum space. %, consider a continuum 2D dimensional system, where the solutions of $\chi(\vec{k})=0$ lie along a circle, which being easily parametrizable allows us to perform the inverse Fourier transform analytically. 
In particular let us consider a model invariant under the shift symmetry $\phi(x)\to \phi(x)+\alpha(x)$ with $\alpha(x)$ satisfying $\mathcal{D}\alpha(x)\equiv (\laplace + 1)\alpha(x)=0$ where $\laplace=\partial_x^2 + \partial_y^2$.
This corresponds to the free part of the 2D UV theory in Ref.~\onlinecite{Lake_2021}, and is invariant under $\alpha_{\vec{k}}(x)= \cos(\vec{k}\cdot \vec{x})$ for any $\vec{k}$ with unit norm ($\|\vec{k}\|=1$) and corresponding conserved quantity $N_{\vec{k}}$.
Since this set can be easily parametrized in polar coordinates, one can exactly compute the inverse Fourier transform. A naive guess to find the most localized quantity in real space is to consider the constant function $f(\vec{k})=1$ with support on this set. The result can be obtained via the line integral
\begin{equation}
	\alpha(\vec{r})=\int_{S^1}\mathrm{d}s \mathop{}\!\mathrm{e}^{\mathrm{i}\vec{k}(s)\cdot \vec{r}}=2\pi J_0(\|\vec{r}\|),
\end{equation}
which leads to a conserved quantity ``localized'' around $\vec{r}=\vec{0}$ with asymptotic behavior (i.e., in the limit $\|\vec{x}\|\gg1$) given by $\alpha(\vec{r})\sim \frac{1}{\sqrt{ \|\vec{r}\| }} \cos(\|\vec{x}\| - \frac{\pi}{4})$.

In fact, the asymptotic behavior is generic and applies even if we consider the inverse Fourier transform of some more generic function along the circle. Let us fix a direction $\vec{r}/\|\vec{r}\|$ and write $\vec{k}\cdot\vec{r} = r \cos{\theta}$ along the unit circle in k-space. Let us take an arbitrary function $f(\theta)$ that we wish to Fourier transform. In the limit $r \to \infty$, we can evaluate this by a stationary phase approximation, which gives
\begin{equation}
    \int_{0}^{2\pi}\mathrm{d}\theta\mathop{}\!f(\theta) \mathrm{e}^{\mathrm{i}r\cos{\theta}} \sim \sqrt{\frac{2\pi}{r}} \Big( \mathrm{e}^{\mathrm{i}(r-\frac{\pi}{4})} f(0) + \mathrm{e}^{-\mathrm{i}(r-\frac{\pi}{4})} f(\pi) \Big) ,
\end{equation}
where we dropped terms that decay faster than $r^{-1/2}$. The leading term might vanish if $f(\theta)$ happens to be zero at both $\theta=0$ and $\pi$; however, for any particular choice of $f$, this will only happen for a few specific directions; in almost all directions we have a decay $r^{-1/2}$ to leading order.

While the circular shape simplified the calculation, this discussion is quite general. What we needed is that for any choice of direction, $\hat{\vec{r}} \equiv \vec{r}/\|\vec{r}\|$
%= (\cos{\theta},\sin{\theta})$ 
there is some isolated points along the loop of conserved momenta where 
%normal to the loop has non-zero overlap with $\hat{\vec{r}}$; 
 $\hat{\vec{r}}$ is normal to the loop;
the integral can then be evaluated in a stationary phase approximation at these points which lead to the same $r^{-1/2}$ decay. 

As an example of how this works, let us consider the model in Eq.~\eqref{eq:Hex} in detail. 
We want to evaluate the line integral 
\begin{equation} \label{eq:linint6}
    \int_{\mathcal{C}}\mathrm{d}s \mathop{}\!\mathrm{e}^{\mathrm{i}\vec{k}(s)\cdot \vec{r}}f(s)=\int_{\mathcal{C}}\mathrm{d}s \mathop{}\!\mathrm{e}^{\mathrm{i}rg(s)}f(s)
\end{equation}
for some function $f$, along the ``loop'' $\mathcal{C}$ within the BZ, defined by the equation $\cos(k_x) + \cos(k_y) - 2 \cos(k_x)\cos(k_y) = 0$, and with $g(s)\equiv \vec{k}(s)\cdot \hat{\vec{r}}$.
As before, the asymptotics will be dominated by saddle points of $g(s)$, i.e., points along the loop where $\hat{\vec{r}} \cdot \vec{k}'(s) = 0$ such that $\hat{\vec{r}}$ is normal to the loop. 
Apart from non-generic behavior along certain direction,  in general, the result decays as $r^{-1/2}$. \

%\begin{equation}\label{eq:IFT}
%    \int_{-\log{3}}^{+\log{3}}\mathrm{d}T\mathop{}\!\frac{f(T)}{J(T)}\mathrm{e}^{\mathrm{i}r g_\theta(T)},
%\end{equation}
% where $J(T)=\sin(k_x(T))\sin(k_y(T))$ is a Jacobi determinant factor and $g_\theta(T) \equiv k_x(T) \cos{\theta} + k_y(T) \sin{\theta}$. 

Apart from their slow asymptotic decay, the real-space conserved quantities constructed from the inverse Fourier transform also exhibit a rich spatial structure, involving short-scale oscillations similar to the circular case considered above.
To address this question we notice that $\mathcal{C}$ splits into four arcs $\mathcal{C}_i$ each of them lying within one of the four quadrants of the BZ, such that Eq.~\eqref{eq:linint6} can be written as
\begin{align} \label{eq:linint62}
     \nonumber &\sum_{i=1}^4\int_{\mathcal{C}_i}\mathrm{d}s \mathop{}\!\mathrm{e}^{\mathrm{i}\vec{k}(s)\cdot \vec{r}}f(s)\\
     &=\int_{\mathcal{C}_1}\mathrm{d}s \mathop{}\!\cos\bigl(k_x(s)x)\cos\bigl(k_y(s)y)f(s).
\end{align}

\begin{figure}[t]
 	\centering
  	 \makebox[\linewidth][c]{\includegraphics[width=\linewidth]{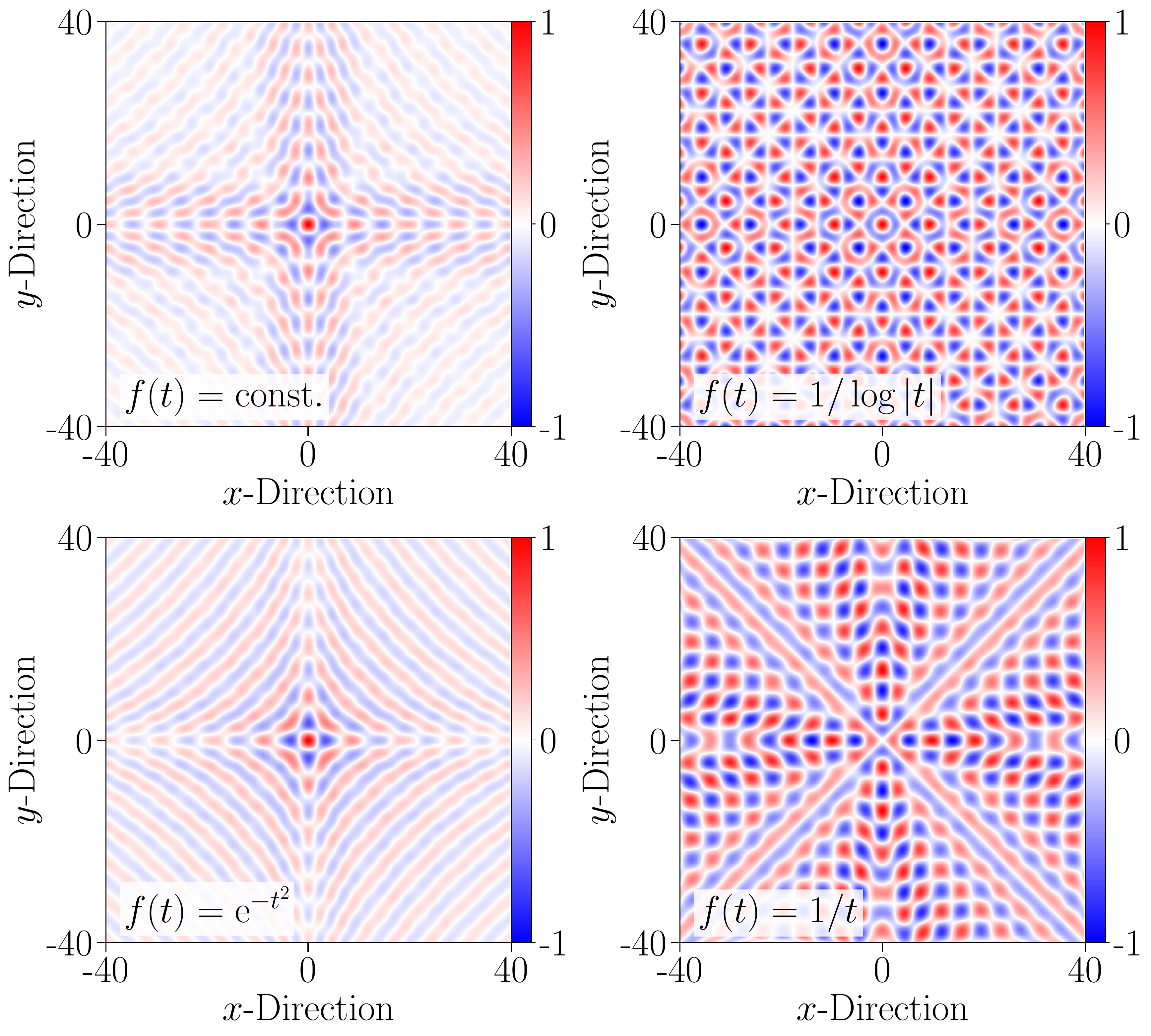}}
  	\caption{Real-space modulations $\alpha(\vec{r})$ of conserved quantities from performing the integral in Eq.~\ref{eq:IFT}.}
 	\label{fig:realspace}
\end{figure}
\newpage

In particular, $\mathcal{C}_1$ lies in the first quadrant with $k_x,k_y>0$, and takes the form of an hyperbola $XY=1$ with $X=1-2\cos(k_x)$, $Y=1-2\cos(k_y)$.
We therefore parametrize it via $X=\mathrm{e}^t$, and $Y=\mathrm{e}^{-t}$, with $t$ taking values in $|t| \leq \log(3)$. In this new parametrization the modulation $\alpha(\vec{r})$ reads
\begin{equation} \label{eq:IFT}
     \int_{-\log(3)}^{\log(3)}\mathrm{d}t\mathop{}\!\|\vec{k}^\prime(t)\|\cos(k_x(t)x)\cos(k_y(t)y)f(t),
\end{equation}
where $\|\vec{k}^\prime(t)\|=\sqrt{(k^\prime_x(t)) ^2+ (k^\prime_y(t)) ^2}$.

In Fig.~\ref{fig:realspace}, we plot a few different results obtained by numerically evaluating the integral~\eqref{eq:IFT} for different choices of $f(t)$. 

Finally, we note that even in cases that evade our stationary phase analysis, the spatial decay of the resulting real-space functions still tends to remain slow. In particular, integrating over a square $[-k,k]\times[-k,k]$ in momentum space gives 
\begin{align*}
\frac{1}{4}\int_{-k}^{k}\text{d}k_x \int_{-k}^{k}&\text{d}k_y\mathop{}\!\mathrm{e}^{\mathrm{i}(k_x x + k_y y)} =  \\ 
&= \frac{\sin(kx)\cos(ky)}{x} + \frac{\sin(ky)\cos(kx)}{y},    
\end{align*}
so that the decay is merely enhanced to $r^{-1}$. 

\end{document}